\documentclass[prd,reprint,nofootinbib]{revtex4-1}
\usepackage{bm}
\usepackage{amssymb,amsmath,graphicx}
\usepackage[colorlinks=true,citecolor=blue]{hyperref}

\newcommand{\rmd}{{\rm d}}    
\newcommand{\rmD}{{\rm D}}

\newcommand{\rmL}{{\rm L}}    
\newcommand{\rmH}{{\rm H}}
\newcommand{\rmA}{{\rm A}}
\newcommand{\rmB}{{\rm B}}
\newcommand{\rmT}{{\rm T}}
\newcommand{\rmCM}{{\rm CM}}

\newcommand{\ItP}{\mathcal{P}}  
\newcommand{\ItF}{\mathcal{F}}
\newcommand{\ItJ}{\mathcal{J}}
\newcommand{\ItQ}{\mathcal{Q}}
\newcommand{\ItZ}{\mathcal{Z}}

\begin{document}
\title{Bobbing and Kicks in Electromagnetism and Gravity}
\author{Samuel E. Gralla}
\author{Abraham I. Harte}
\author{Robert M. Wald}
\affiliation{\it Enrico
Fermi Institute and Department of Physics, University of Chicago
\\ \it 5640 South Ellis Avenue, Chicago, Illinois 60637, USA}

\begin{abstract}

We study systems analogous to binary black holes with spin in order to
gain some insight into the origin and nature of ``bobbing'' motion and
``kicks'' that occur in this system. Our basic tool is a
general formalism for describing the motion of extended test bodies in
an external electromagnetic field in curved spacetime and possibly
subject to other forces.  We first show that bobbing of exactly the
type as observed in numerical simulations of the binary black hole system
occurs in a simple system consisting of two spinning balls connected
by an elastic band in flat spacetime. This bobbing
may be understood as arising from the difference between a spinning
body's ``lab frame centroid'' and its true center of mass, and is
purely ``kinematical'' in the sense that it will appear regardless of
the forces holding two spinning bodies in orbit. Next, we develop
precise rules for relating the motion of charged bodies in a
stationary external electromagnetic field in flat spacetime with the
motion of bodies in a weakly curved stationary spacetime. We then
consider the system consisting of two orbiting charges with magnetic
dipole moment and spin at a level of approximation corresponding to
1.5 post-Newtonian order. Here we find that considerable amounts of
momentum are exchanged between the bodies and the electromagnetic
field; however, the bodies store this momentum entirely as ``hidden''
mechanical momentum, so that the interchange does {\em not} give rise
to any net bobbing. The net bobbing that does occur is due solely to
the kinematical spin effect, and we therefore argue that the net
bobbing of the electromagnetic binary is not associated with possible
kicks.  We believe that this conclusion holds in the
gravitational case as well.

\end{abstract}


\maketitle

\section{Introduction}

The successful simulation of binary black holes with spin  \cite{Rezzolla, Campanelli07, Campanelli07-2, Gonzalez07} has presented two intriguing surprises: If the spin lies in the
orbital plane, an up-and-down ``bobbing'' of the entire orbital plane
is seen to occur in phase with the orbit, and when the black holes
merge, the final black hole can acquire a ``kick'' velocity
exceeding those occurring in the non-spinning case by an order of
magnitude.  In the kick-maximizing case of circular orbit with
antiparallel spins lying in the orbital plane, the direction (up or
down) and speed of the kick depend on the orbital phase at merger.
Two natural questions are: What causes the bobbing motion?  Can the kick
be viewed as a post-merger continuation of the bobbing?

In this paper we shall address these questions by studying simple
\textit{analogous systems}: analytical toy models of the spinning
black hole binary system that display some of its salient features.
We first consider two spinning bodies connected by an elastic band
in flat spacetime, a system we refer to as ``tetherballs.''  We find that this system displays bobbing motion very similar to that observed in numerical simulations of binary black holes. The bobbing is identified as arising from a purely ``kinematical'' spin effect that gives rise to bobbing independently of the details of the particular type
of force holding the two bodies in orbit.  This effect is analogous to the Thomas precession, which gives rise to a precession independently of the details of the particular type of force accelerating a spinning body.

The analog system of primary interest to us is an \textit{electromagnetic binary}, consisting of two charged magnetic dipoles with spin in flat spacetime.  Here we find that, in addition to the ``kinematical bobbing'' associated with their spin, the individual bodies also undergo ``dynamical bobbing'' associated with their magnetic dipole moment. However, at the level of approximation corresponding to 1.5 post-Newtonian---which we refer to as ``1.5 post Coulombic''---we shall show that the mass-weighted dynamical bobbing of the two bodies is equal and opposite, so that the {\it net} bobbing (i.e., the sum of the mass times the bobbing velocity for the two bodies) is purely kinematical, just as in the case of tetherballs. Nevertheless, we will also see that a considerable amount of momentum is exchanged between the bodies and the electromagnetic field during each orbit. This momentum exchange does not contribute to the net bobbing because the net momentum acquired by the bodies from the electromagnetic field goes into the ``hidden mechanical momentum'' \cite{Jackson, griffiths} of the bodies.

If an electromagnetic analog of a merger event were to occur, a net kick could be experienced by the system only if momentum is carried away to infinity by the electromagnetic field. We cannot analyze the momentum radiated to infinity---momentum cannot be radiated to infinity at the 1.5 post-Coulombic approximation---but, in a more exact treatment, we would expect the radiated momentum to be related to the total momentum stored in the electromagnetic field at or near merger. However, the momentum stored in the electromagnetic field is entirely unrelated to the net bobbing of
the bodies. In particular, the momentum stored in the electromagnetic
field depends upon the magnetic dipole moment of the bodies but not
their spin, whereas the net bobbing depends upon their spin
but not their magnetic dipole moment. Thus, in the electromagnetic case, we believe that the bobbing phenomenon is entirely unrelated to possible kicks.

In this paper, we will derive
precise rules for relating the equations of motion of charged bodies in a
stationary external electromagnetic field in flat spacetime with the
motion of bodies in a weakly curved stationary spacetime. These rules will
enable us to directly translate our electromagnetic results on bobbing motion into
results for the corresponding gravitational case---i.e., a binary system of spinning masses---at the 1.5
post-Newtonian approximation. This gravitational binary system has been directly analyzed in the
1.5 post-Newtonian approximation in \cite{ThorneBobbing}, and our results on bobbing in the gravitational case agree with theirs. However, it is much more difficult to unambiguously interpret the results in the gravitational case as compared with the electromagnetic case for the following three reasons:
(i) In the gravitational case,
the kinematical bobbing terms take the exactly same mathematical form as the dynamical
bobbing terms, so it is much less clear how to make a clean distinction between effects
of a kinematical origin and effects of a dynamical origin.  (ii) The description of
the motion of bodies in the gravitational
case is gauge dependent, as are the ``gravitational field strengths.''
(iii) In general relativity, only the total momentum of the entire system is well defined. Any allocation of a portion of the
momentum of the system to the gravitational field and a portion to the bodies requires both
a choice of pseudotensor (such as the Landau-Lifshitz pseudotensor
used in \cite{ThorneBobbing}) and a choice of gauge, and it is unclear the extent to which this allocation
may depend on these choices.

Nevertheless, the very close analogy established in this paper between the
electromagnetic and gravitational cases leads us to believe that our interpretation of the nature of bobbing and kicks in the electromagnetic case should extend to the gravitational case. We therefore
propose the following answers to the
questions posed in the first paragraph above: The bobbing seen in
simulations of black hole mergers is a purely kinematical spin effect
that is in no way special to the gravitational interaction. The kick
has an entirely different cause, and should not be viewed as a
post-merger continuation of the bobbing phenomena\footnote{As previously mentioned, in the electromagnetic case the net bobbing depends upon spin of the bodies, whereas the momentum stored in the electromagnetic field (which we have associated with kicks) depends upon magnetic dipole moment of the bodies. Under the translation rules between electromagnetism and gravity established in section  \ref{Sect: Analogy}, these distinct terms in electromagnetism translate into terms of exactly the same functional form in gravity. Therefore, it should not be surprising if in the gravitational case, the magnitude of the kick velocity and its
dependence on the phase of the orbit are very similar to those of
the net bobbing velocity.}.

The nature of the effects we wish to analyze in this paper require us to be very careful and precise in our definitions of quantities such as the linear momentum, spin, and center of mass position of a body, which will require us to develop a substantial amount of machinery in the next section. It may therefore be worthwhile to give here a simple
explanation of the nature and origin of the kinematical bobbing effect in flat spacetime.  Consider a body described by stress-energy tensor $T_{\mu \nu}$ in global inertial coordinates $(t,x^i)$ that is acted upon by an
external $4$-force density $f^\mu$, so that $T_{\mu \nu}$ satisfies
\begin{equation}\label{ex:diff}
\partial_\mu T^{\mu \nu} = f^\nu.
\end{equation}
Integrating this equation, we obtain
\begin{equation}\label{ex:int}
\frac{\rmd}{\rmd t} \int T^{0\nu} \rmd^3 x = \int f^\nu \rmd^3 x.
\end{equation}
Multiplying the $\nu=0$ component of Eq. \eqref{ex:diff}
by $x^i$, integrating over space, integrating by parts, and using \eqref{ex:int}, we obtain
\begin{equation}
{\mathcal E} \frac{ \rmd z^i_{\rm L}}{\rmd t} = \int T^{0i} \rmd^3 x + \int f^0(x^i-z_{\rm L}^i)\rmd^3 x .
\label{eomzl}
\end{equation}
where ${\mathcal E} =\int T^{00} \rmd^3 x$ is the total energy of the body as measured in the lab frame, and we have defined the ``lab frame
centroid''
\begin{equation}
z^i_{\rm L} \equiv \frac{1}{\mathcal E} \int T^{00} x^i \rmd^3 x .
\end{equation}
Differentiating \eqref{eomzl} with respect to time
and using Eq. \eqref{ex:int} again, we obtain
\begin{align}
{\mathcal E} \frac{ \rmd^2 z^i_{\rm L}}{\rmd t^2} = \int f^i \rmd^3 x + \frac{\rmd}{\rmd t} \int f^0(x^i-z_{\rm L}^i)\rmd^3 x \nonumber
\\
~ - \frac{\rmd z_{\rm L}^i}{\rmd t} \int f^0 \rmd^3 x .
\label{ex:lab}
\end{align}

\begin{figure}[t]
\centering
\includegraphics[width=3 in]{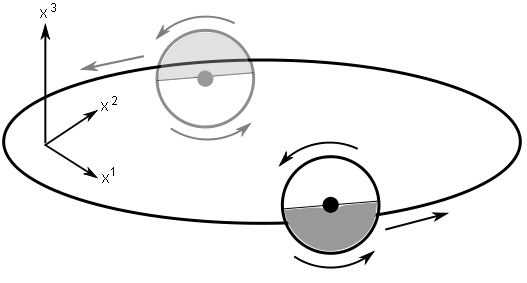}
\caption{An Illustration of kinematical bobbing.  The lab frame centroid of a body (solid line) is confined to the $x^1$-$x^2$ plane.  However, since the body has a (constant) spin vector lying in the plane, its center of mass (solid dot) is displaced in the $x^3$ direction by a velocity dependent factor (equation \eqref{ex:labrest}), giving rise to bobbing.  The body is shown at two phases in its orbit.  In the foreground, the bottom portion of the body contains extra lab frame kinetic energy (illustrated with shading), skewing the lab frame centroid downwards so that the center of mass must have bobbed up.  One half of an orbit later (shown in the background), the situation is reversed and the center of mass has bobbed down.  If two bodies with anti-aligned spins were considered, a net bobbing would be observed.}\label{fig:kbobbing}
\end{figure}

We now restrict consideration to
non-relativistic motion and drop terms of
order $ | \rmd \vec{z}_{\rm L} / \rmd t |^2$.  We may then replace $\mathcal E$ by the rest mass\footnote{The rest mass of a body is
given by $m = (-p^\mu p_\mu)^{1/2}$, where $p^\mu$ is the $4$-momentum of the body, defined (in the absence of an electromagnetic field) by an integral of $T^{\mu \nu}$ over a spacelike slice. However, even in Minkowski spacetime, if an external force acts
on a  body, then $p^\mu$ will be slice dependent. In particular, $p^\mu$ and $m$ depend on the choice of frame and we must distinguish between, e.g., the rest mass $m_{\rm L}$ as determined in the lab frame and
the rest mass $m_{\rm CM}$ as determined in the center of mass frame. These differences will be negligible for the present calculation, and we will ignore them. Similar remarks apply to spin. We will treat these issues with great care in section \ref{Sect: Mechanics}.}, $m$, of the body.
If the integrated $4$-force density is nearly orthogonal
to the $4$-velocity of the lab frame centroid (as must occur if the rest mass of the body is approximately constant), then the last term on the right side of \eqref{ex:lab} will be of order $| \rmd \vec{z}_{\rm L}/\rmd t |^2$ and may be dropped.
Furthermore, if the external force is nearly centered about
the lab frame centroid, the second term on the right side of
\eqref{ex:lab} also will be negligible\footnote{If this term is not
  negligible, it may contribute to additional ``dynamical
  bobbing'' of the body. This occurs in the electromagnetic case.}. Dropping this term as well, we obtain
\begin{equation}
m \frac{\rmd^2 z^i_{\rm L}}{\rmd t^2} \simeq \int f^i \rmd^3 x ,
\label{Fma}
\end{equation}
which can be recognized as a version of $\vec{F} = m
\vec{a}$. However, the ``$\vec{a}$'' that appears in this equation is the acceleration of the lab frame centroid. The lab frame centroid $z^i_{\rm L}$ measures the ``center of energy'' of a body as determined by a lab frame observer.  However, for a spinning, moving body, this center of energy will be skewed towards the side of the (rotating) body that moves with the body's bulk motion, on account of the extra lab frame kinetic energy present there.  As obtained in equation \eqref{DeltaZ2} below, the true center of mass $z^i_{\rm CM}$ is related to the lab frame centroid $z^i_{\rm L}$ approximately by\footnote{A derivation of equation \eqref{ex:labrest} is also given in \cite{ThorneBobbing} (see also \cite{MTW,DixonSR}). However, the notion of center of mass used in \cite{ThorneBobbing} corresponds to the ``hybrid'' definition of center of mass introduced in subsection \ref{Sect: CM} and does not agree with the notion of center of mass used in the present paper.}
\begin{equation}\label{ex:labrest}
m \vec{z}_{\rm CM} = m \vec{z}_{\rm L} + \vec{S} \times \frac{\rmd \vec{z}_{\rm L}}{\rmd t} + O(| \frac{\rmd \vec{z}_{\rm L}}{\rmd t} |^2).
\end{equation}
If $\int f^i \rmd^3 x$ lies in a plane at all times, and if
$\rmd z^i_{\rm L}/\rmd t$ initially lies in this plane, then by \eqref{Fma} the lab frame centroid is confined to the plane. However, if the spin of the body has a component in the plane, then by \eqref{ex:labrest} the center of mass will be displaced out of the plane.  This accounts for the ``kinematical bobbing'' (see figure \ref{fig:kbobbing}).

The remainder of this paper is organized as follows. In section \ref{Sect: Mechanics} we present a general formalism for describing the motion of extended test bodies in curved spacetime.  We explicitly include the possibility of an external electromagnetic field, and allow for the possibility of additional external forces.  The first subsection defines linear and angular momenta of a body.  The second subsection deals with the association of a center of mass worldline to a body.  The third subsection deals with the evaluation of electromagnetic and gravitational forces and torques up to quadrupole order. Section \ref{Sect: Mechanics} is almost entirely review, although some aspects of the presentation are new and there is some new material on, e.g., the relationship between the true center of mass and the ``lab frame'' and ``hybrid'' centroids.

In section \ref{Sect: SimpleBobbing} we apply this formalism to derive bobbing effects in simple example systems.  First we consider the case of tetherballs and derive the kinematical bobbing effect discussed above.  We then consider the example of a magnetic dipole orbiting a line electric charge in flat spacetime, and a spinning body orbiting a line mass in general relativity. In these cases the kinematical bobbing is supplemented by dynamical bobbing.

In section \ref{Sect: Analogy} we give a careful treatment of the analogy between the equations of motion of a body moving in the electromagnetic field of a stationary source in flat spacetime and the motion of a body in a weakly curved, stationary spacetime in general relativity. We extend the results of \cite{WaldSpin} to include quadrupole terms and terms quadratic in the velocity of the body. We provide precise rules for translating the electromagnetic evolution equations for center of mass motion and spin into the corresponding gravitational equations at this order of approximation.

Finally, in section \ref{Sect: GeneralEM} we consider the
electromagnetic analog of a black hole binary, consisting of two orbiting charged magnetic dipoles with spin.  We assume that each body moves in the electromagnetic field of the other according to the laws of test body motion derived in section \ref{Sect: Mechanics}.  (These laws were also obtained in \cite{HarteEM, GrallaHarteWald}, where self-force effects were considered in addition.)  We obtain the spin and magnetic dipole terms appearing in the equations of motion at ``1.5 post-Coulombic'' order (corresponding to 1.5 post-Newtonian order in the gravitational case). Using the translation rules developed in section \ref{Sect: Analogy} we also obtain the corresponding results for gravity, which agree with \cite{ThorneBobbing}.  We analyze the flow of momentum between the bodies and the electromagnetic field, and discuss what can be learned from this analysis about kicks.

Our notation and conventions on indices are as follows. Latin indices from early in the alphabet ($a,b,c,\dots$) are used to denote abstract tensor indices. Greek indices are used to denote components of these tensors (most commonly in global inertial coordinates in Minkowski spacetime). Latin indices from the middle of the alphabet ($i,j,k,\dots$) denote spatial components, and a ``$0$'' will be used for time
components. Sign conventions are those of \cite{Wald}.

\section{Mechanics of extended test bodies}
\label{Sect: Mechanics}

Consider a body with stress-energy tensor $T^{ab} = T^{(ab)}$ and  electromagnetic charge-current density $J^a$ moving in a curved spacetime $(M,g_{ab})$. Both $T^{ab}$ and $J^a$ are assumed to have support within a spatially-compact worldtube $W$. We also allow for an electromagnetic field $F_{ab}$, which is assumed to satisfy Maxwell's equations
\begin{subequations}
\label{Maxwell}
\begin{align}
  \nabla_{[a} F_{bc]} =0
  \\
  \nabla^b F_{ab} = 4 \pi J_a
\end{align}
\end{subequations}
in a neighborhood of $W$. It follows from this that the current density is conserved: $\nabla_a J^a =0$.

The total stress-energy tensor $T_{\mathrm{tot}}^{ab}$ of the system is also assumed to be conserved. (This would follow from Einstein's equation, but we do not impose Einstein's equation in the present section.) $T_{\mathrm{tot}}^{ab}$ is the sum of the body stress-energy $T^{ab}$, the stress-energy of the electromagnetic field
\begin{equation}
  T^{ab}_{\mathrm{EM}} = \frac{1}{4 \pi} (F^{ac} F^{b}{}_{c} - \frac{1}{4} g^{ab} F_{cd} F^{cd}),
\label{TEM}
\end{equation}
and the stress-energy $T^{ab}_{\mathrm{ext}}$  associated with any other fields or bodies that may be present. Conservation of this total yields
\begin{equation}
  \nabla^b T_{ab} = F_{ab} J^b + f_a ,
  \label{StressCons}
\end{equation}
where $F^{ab} J_b  = - \nabla_b  T^{ab}_{\mathrm{EM}} $ is the ordinary Lorentz force density and $f^a \equiv - \nabla_b  T^{ab}_{\mathrm{ext}} $ is the ``external force density.'' Except in Sect. \ref{Sect: tether} below, this paper will mainly be concerned with the case $f_a=0$.

We now proceed to review some basic results on the mechanics of compact bodies satisfying the equations just described. Linear and angular momenta are defined in Sect. \ref{Sect: MomDef}, and various aspects of their evolution are discussed there. The notion of a center of mass worldline is then introduced in Sect. \ref{Sect: CM}. For stationary spacetimes, we also define the notion of a lab frame centroid and compare it with the center of mass. Multipole approximations for the force and torque are presented in Sect. \ref{Sect: Forces} up to quadrupole order. The formalism described here is mainly due to Dixon \cite{Dix70a, Dix70b, Dix74} (see also \cite{Dix67, Dix79, Ehl77, HarteThesis, Dix08}), although many elements of our presentation arose more recently in \cite{HarteSyms}.

In order to be mathematically well-defined, the definition of momentum given in Sect. \ref{Sect: MomDef} requires only that $W$ lie in a suitable normal neighborhood of the timelike worldline used in the construction of certain ``generalized Killing fields.'' This imposes relatively weak restrictions on the size of the object and the curvature within it. It does not, however, restrict one to the test body regime. Electromagnetic self-fields may be large, and, if Einstein's equation is imposed, the body can contribute significantly to the spacetime curvature. Some relatively minor additional restrictions are required in Sect. \ref{Sect: CM} in order to ensure the existence of a unique center of mass worldline \cite{Schattner}, but the constructions and results presented there also do not require that any self-fields be negligible.

This contrasts with the multipole approximations given in Sect. \ref{Sect: Forces}, which are valid in the form presented only for sufficiently small test bodies. Despite this, it can be shown that the definition of momentum given in \eqref{PDef} below can be appropriately modified so that essentially identical approximations remain correct even in the presence of large self-fields \cite{HarteScalar, HarteEM, HarteHighMoments}. Indeed, using a modified definition of the quadrupole and higher moments of $T_{ab}$ (but not the moments of $J^a$), the laws of motion are changed only in that each instance of the external field is replaced by that external field plus the so-called ``regular'' Detweiler-Whiting component of the body's self-field \cite{DetWhiting}. A short review of the full theory may be found in \cite{HarteHighMoments} for the special case of self-interaction with a scalar field in a fixed curved spacetime. This generality will not be needed here, so we shall adopt definitions equivalent to Dixon's for simplicity. Consequently, the applicability of the results of this section should be considered as restricted to the test body regime.

\subsection{Momentum}
\label{Sect: MomDef}

Standard definitions for a body's linear and angular momenta in flat spacetime may be found in textbooks (see, e.g., \cite{MTW, DixonSR}). In global inertial coordinates $x^\mu$, they are given by
\begin{subequations}
\label{MinkMomenta}
\begin{align}
  p^\mu &= \int_\Sigma T^{\mu\nu} \rmd S_\nu
  \\
  S^{\mu \nu } (z) = S^{[\mu \nu]} (z) &= 2 \int_\Sigma (x-z)^{[\mu} T^{\nu]\lambda} \rmd S_\lambda,
  \label{FlatS}
\end{align}
\end{subequations}
where $z^\mu$ is an arbitrarily chosen origin in spacetime and $\Sigma$
any Cauchy hypersurface. It is common to also introduce a spin angular momentum vector
\begin{equation}
  S_\mu = - \frac{1}{2} \epsilon_{\mu\nu\lambda\rho} \left( \frac{p^\nu}{\sqrt{-p_\kappa p^\kappa}} \right) S^{\lambda \rho}(z).
\end{equation}
Assuming that $F_{ab} J^b + f_a = 0$, it immediately follows from \eqref{StressCons} that $p^\mu$, $S^{\mu \nu}$, and $S_\mu$ do not depend on $\Sigma$ for systems with finite radius. $p^\mu$ and $S_\mu$ are independent of $z^\mu$ as well.

It is useful to rewrite \eqref{MinkMomenta} without any reference to particular coordinate systems. The resulting procedure combines both the linear and angular momenta into a single expression, and makes explicit that they are
both associated with Poincar\'{e} symmetries. Given any Killing field
$\xi^a$, let
\begin{equation}
  \ItP_\xi = \int_\Sigma T^{a}{}_{b} \xi^b \rmd S_a.
  \label{PXiSR}
\end{equation}
$\ItP_\xi$ may be viewed as a linear map from the space of Killing fields into $\mathbb{R}$. Since any Killing field is uniquely characterized by its value and that of its first derivative at an arbitrarily chosen point $z$, we may express $ \ItP_\xi$ as a linear map on the space of ``Killing data" $\xi^a(z)$ and $\nabla_a \xi^b(z)$. It therefore has the form
\begin{equation}
 \ItP_\xi = p^a(z) \xi_a(z) + \frac{1}{2} S^{ab}(z) \nabla_a \xi_b(z)
  \label{pToPSR}
\end{equation}
for some tensors $p^a(z)$ and $S^{ab} = S^{[ab]}(z)$ defined at $z$. These are exactly the linear and angular momenta given by the more familiar definition \eqref{MinkMomenta}. Note that for fixed $\xi^a$, $\ItP_\xi$ is independent of both $z$ and $\Sigma$ when $F_{ab} J^b + f_a =0$. It therefore generates a class of conserved scalars in this case.

Eqs. \eqref{PXiSR} and \eqref{pToPSR} are trivially extended to apply in any maximally symmetric spacetime. As we shall now show, a suitable generalization of the notion of a Killing field allows them to be applied in generic spacetimes \cite{HarteSyms}.

\subsubsection{Generalized Killing fields}
\label{Sect: GKFs}

The appropriate notion of a generalized Killing field will depend upon the choice of a smooth timelike worldline $\ItZ = \{ z(s) | \forall s \in \mathbb{R} \}$ and a unit future-directed timelike vector field $n^a(s) \in T_{z(s)} M$ defined along $\ItZ$. Note that $s$ is not required to be proper time and $n^a$ need not lie tangent to $\ItZ$. A ``center of mass frame'' will be defined in Sect. \ref{Sect: CM} below that uniquely fixes $\ItZ$ and $n^a(s)$. For the constructions and definitions of this subsection, however, they may be chosen arbitrarily, subject only to the body being contained in the normal neighborhood of $\ItZ$ defined below.

A specific generalized Killing field may be fixed by choosing a time $\bar{s}$ together with tensors $\mathcal{A}_a(\bar{s})$ and $\mathcal{B}_{ab} = \mathcal{B}_{[ab]}(\bar{s})$ at $z(\bar{s})$.  We use the Killing transport equations
\begin{align}
    & \frac{\rmD }{ \rmd s } \mathcal{A}_a(s) - \dot{z}^b(s) \mathcal{B}_{b a}(s) = 0
    \label{KT1}
    \\
    & \frac{\rmD }{ \rmd s } \mathcal{B}_{ab}(s)+ R_{a b c}{}^{d}(z(s)) \dot{z}^c(s) \mathcal{A}_d(s) = 0
    \label{KT2}
\end{align}
to uniquely extend these tensors to all of $\ItZ$, where
\begin{equation}
 \frac{\rmD }{ \rmd s } \equiv \dot{z}^a \nabla_a .
\end{equation}
Note that the skew symmetry of $\mathcal{B}_{ab}$ is preserved by this prescription\footnote{It is possible to use initial data for which $\mathcal{B}_{ab}$ is not symmetric. The vector field that eventually results generalizes a homothety or other non-Killing affine collineation \cite{HarteSyms}.}.

Now consider all pairs $(z,v^a)$, where $z\in \ItZ$ and $v^a \in T_z M$ is orthogonal to $n^a$. This forms a subset $T_\bot \ItZ$ of the tangent bundle $TM$. For any element of $T_\bot \ItZ$, one may associate an affinely-parameterized geodesic $y(u)$ whose initial point is $y(0) = z$, and whose initial tangent is $v^a$. As long as these geodesics can be extended sufficiently far, the map $(z,v^a) \rightarrow y(1)$ will be a smooth function from $T_\bot \ItZ$ to $M$. Its Jacobian is clearly invertible at every point $(z,0)$, so it follows from the inverse function theorem that the given map defines a diffeomorphism on some neighborhood $\mathcal U \subseteq M$ of $\ItZ$. This will be the region in which the generalized Killing fields (GKFs) are to be defined.

We now define the GKF $\xi_a(x)$ in $\mathcal U$ associated with our choice of initial data $\mathcal{A}_a (\bar{s})$ and $\mathcal{B}_{ab}(\bar{s})$. Let $x \in \mathcal U$. The diffeomorphism described in the previous paragraph may be used to uniquely associate $x$ with an element $(z(\tau_x), v^a(\tau_x))$ of $T_\bot \ItZ$. Use this pair to construct a geodesic $y(u)$ as before. The GKF may then be computed along $y(u)$ by solving the Jacobi (or geodesic deviation) equation
\begin{equation}
  \frac{\mathrm{D}^2 \xi_a}{\rmd u^2} - R_{abc}{}^{d}  \dot{y}^b \dot{y}^c \xi_d = 0 \label{Jacobi}
\end{equation}
with initial data
\begin{align}
  \xi_a (z(\tau_x)) &= \mathcal{A}_a (\tau_x)
  \label{ADef}
  \\
  \frac{\rmD \xi_a(z(\tau_x))}{\rmd u}  &= v^b \mathcal{B}_{ba} (\tau_x).
  \label{BDot}
\end{align}
Here, $\dot{y}^a$ denotes the tangent to $y(u)$. The given equations uniquely define $\xi_a(x)$ throughout $\mathcal U$ once $\mathcal{A}_a$ and $\mathcal{B}_{ab}$ are given at any one point on $\ItZ$.

It will often be useful to introduce the spacelike hypersurface $\Sigma(s) \subset \mathcal U$ as the image of $(z(s),v^a)$ as $v^a$ is varied while $s$ is held fixed (i.e. the union of all geodesics in $\mathcal U$ passing through $z(s)$ and orthogonal to $n^a(s)$ at that point). Restricted to any one of these hypersurfaces, $\xi_a$ is an ordinary Jacobi field. All solutions to the Jacobi equation can be written down explicitly in terms of the first and second derivatives of Synge's world function as well as the initial data $\mathcal{A}_a$ and $\mathcal{B}_{ab}$ \cite{Dix70a}. The first of these two tensors is easily interpreted using \eqref{ADef}: It specifies $\xi_a$ on $\ItZ$. Combining \eqref{KT1} and \eqref{BDot} shows that $\mathcal{B}_{ab}$ serves to fix
\begin{equation}
  \nabla_a \xi_b(z(s)) = \mathcal{B}_{ab} (s)
  \label{BDef}
\end{equation}
for all $s$.

One way to motivate the given definition of the GKFs is to rewrite \eqref{Jacobi} as a pair of first order differential equations. The result is very similar to the Killing transport equations: An analog of \eqref{KT1} carries over exactly, while \eqref{KT2} is modified only by an overall contraction with $\dot{y}^b$. It is necessary to modify the Killing transport equations in this way because inconsistencies might arise if one attempts to propagate derivatives transverse to the direction of transport along all spatial geodesics emanating from a single point. Further details and alternative motivations for our definition of generalized Killing fields may be found in \cite{HarteSyms}.

The space $V(\ItZ, n)$ of all possible GKFs associated with a particular choice of $\ItZ$ and $n^a(s)$ is obtained by varying the initial data $\mathcal{A}_a (\bar{s})$ and $\mathcal{B}_{ab}(\bar{s})$. A general GKF is clearly linear in this data, so $V(\ItZ, n)$ is a vector space. In four spacetime dimensions, $\dim \, V(\ItZ, n) = 10$. Also note that specifying a GKF and its first derivative at any one point on $\ItZ$ fixes it throughout $\mathcal{U}$. Since any true Killing field satisfies both the Killing transport equations along any curve and the geodesic deviation equation along any geodesic, it is clear that true Killing fields are also generalized Killing fields. In a maximally symmetric spacetime, $V(\ItZ, n)$ is exactly the space of all Killing fields (and therefore independent of $\ItZ$ or $n^a(s)$ in such a case). Finally, we note that for any $\xi^a \in V(\ItZ,n)$ and $z \in \ItZ$, one may show that
\begin{equation}
  \mathcal{L}_\xi g_{ab}(z) = \nabla_a \mathcal{L}_\xi g_{bc} (z) =0.
  \label{AlmostKilling}
\end{equation}
From this, it follows by the same derivation as for true Killing fields (see, e.g., \cite{Wald}) that
\begin{equation}
  \nabla_a \nabla_b \xi^c(z) = \xi^d(z) R_{dab}{}^{c}(z) .
  \label{D2Xi}
\end{equation}
for any $z \in \ItZ$.

\subsubsection{Definition of momentum}

We may now define the momentum of an essentially arbitrary compact object moving in a
general spacetime. Choose a timelike worldline $\ItZ$ and a vector field $n^a$ on $\ItZ$ as above. We require only that the body's worldtube $W$ be contained
within the neighborhood $\mathcal{U}$ in which generalized
Killing fields are defined (at least for the time interval under
consideration). The definition \eqref{PXiSR} for the momentum could now be carried over by
simply replacing the arbitrary Killing field appearing there with an arbitrary GKF. It is, however, useful to add an additional term in the presence of an electromagnetic field \cite{Dix70a, Dix74, BaileyIsrael}. For the remainder of this paper, we define $\ItP_\xi$ at time
$s$ by
\begin{align}
  \ItP_\xi  (s) = \int_{\Sigma(s)}  \rmd S_a \bigg[ T^{a}{}_{b}(x) \xi^b (x) + J^a(x) \nonumber
  \\
  \times \int_0^1 \rmd u \, u^{-1} \sigma^{b'} (y',z) \xi^{c'} (y') F_{b'c'}(y') \bigg] .
  \label{PDef}
\end{align}
The $\xi^a$ appearing here is an arbitrary generalized Killing field. $\Sigma(s)$ is the aforementioned spacelike hypersurface generated by geodesics orthogonal to $n^a(s)$ at $z(s)$. The $u$-integral is taken along the affinely-parameterized geodesic $y'(u)$ satisfying $y'(0) = z(s)$ and $y'(1) = x$. Also note that the biscalar $\sigma$ is Synge's world function. Its derivative $\sigma_{a'}(y',z) = \nabla_{a'} \sigma(y',z)$ is well-known to lie tangent to $y'(u)$.

The additional electromagnetic term appearing in \eqref{PDef} as compared with \eqref{PXiSR}
is useful for constructing conservation laws in symmetric systems: See \cite{Dix70a} or Eq. \eqref{ConsLaw} below. It also arises from the general theory of multipole moments describing $J^a$ and $T_{ab}$. At least if $f_a = 0$, there is a sense in which evolution equations for $\ItP_\xi$ as defined above are the only ones implied by stress-energy conservation \cite{Dix74}. The associated linear and angular momenta---see \eqref{pToP} below---are also the canonical choices in a Lagrangian formalism describing the motion of charged particles with respect to some reference worldline (such as $\ItZ$) \cite{BaileyIsrael}. However, in a multipole expansion, the dominant contribution of this additional electromagnetic term will enter via the electric dipole moment of the body. Since we will be interested below only in the case of bodies with vanishing electric dipole moment,
this term will play essentially no role in the remainder of this paper. Nevertheless, we include it here
for completeness.

Since $\ItP_\xi(s)$ is linear in $\xi^a$, and each GKF is uniquely determined by its value and that of its derivative at $z(s)$, we may write
\begin{equation}
  \ItP_\xi(s) = p^a \xi_a(z(s)) + \frac{1}{2} S^{ab} \nabla_a \xi_b(z(s))
  \label{pToP}
\end{equation}
for some $p^a$ and $S^{ab} = S^{[ab]}$. It is not difficult to see that these tensors depend only on the single point $z(s)$ and vector $n^a(s)$; the extension of $\ItZ$ and $n^a$ away from $z(s)$  does not affect any GKF restricted to $\Sigma(s)$, and is therefore irrelevant for the definitions of $p^a$ and $S^{ab}$ at $s$. Although presented differently, the definitions \eqref{PDef} and \eqref{pToP} for $p_a(z,n)$ and $S_{ab}(z,n)$ are equivalent to the ones originally given by Dixon \cite{Dix70a, Dix74}.

We now derive equations for the time evolution of the momenta. For this purpose, it is convenient to rewrite \eqref{PDef} in terms of a vector potential $A_a$ satisfying $F_{ab} = 2 \nabla_{[a} A_{b]}$. Integrating by parts, we obtain
\begin{align}
  & \ItP_\xi(s) = \int_{\Sigma(s)} \rmd S_a \bigg\{ T^{a}{}_{b}(x) \xi^b(x) + J^a(x)
  \nonumber
  \\
  & ~ \times \bigg[(A_b \xi^b)|_z^x -  \int_0^1 \rmd u \, u^{-1} \sigma^b(y,z) \mathcal{L}_\xi A_b(y) \bigg] \bigg\} .
  \label{PDefwA}
\end{align}
Differentiating this using \eqref{StressCons} yields
\begin{align}
  \frac{\rmd}{\rmd s} ( \ItP_\xi + q A_a \xi^a ) = \int_\Sigma \rmd S_c w^c \bigg[ \frac{1}{2} T^{ab} \mathcal{L}_\xi g_{ab} + f_a \xi^a \nonumber
  \\
  ~ + J^a \bigg( \mathcal{L}_\xi A_a - \nabla_a \int_0^1 \rmd u \, u^{-1} \sigma^b \mathcal{L}_\xi A_b \bigg) \bigg] ,
  \label{PDot}
\end{align}
where $w^a$ is the time evolution vector field for the hypersurfaces
$\Sigma(s)$. The combination $A_a \xi^a$ on the left-hand side is understood to be
evaluated at $z(s)$, and $q$ is the total electric charge
\begin{equation}
  q = \int_\Sigma J^a \rmd S_a .
\end{equation}

One important special case of \eqref{PDot} arises in the presence of an exact Killing vector $\zeta^a$ that also preserves the electromagnetic field, i.e., $\mathcal{L}_\zeta g_{ab} = \mathcal{L}_\zeta F_{ab} = 0$. It is then possible to choose an $A_a$ such that $\mathcal{L}_\zeta A_a =0$. In such a gauge, it is clear that the right-hand side of \eqref{PDot} vanishes when $f_a =0$. This means that
\begin{equation}
  \ItP_\zeta(s) + q (\zeta^a A_a )|_{z(s)}
  \label{ConsLaw}
\end{equation}
is a constant of the motion. In the absence of an electromagnetic field (or if $q=0$), $\ItP_\zeta$ is conserved by itself.

Another interesting application of  the constancy of \eqref{ConsLaw} is to the case of a charged particle in flat spacetime with an electromagnetic field satisfying $\nabla_a F_{bc}=0$. If $\zeta^a$ is a purely translational Killing field (so $\nabla_a \zeta^b =0$), one finds that
\begin{equation}
  \frac{\rmd \ItP_\zeta}{\rmd s} = \zeta^a \frac{\rmD p_a}{\rmd s}  = q \zeta^a F_{ab} \dot{z}^b.
\end{equation}
This has the form of the Lorentz force law, and is valid without any restriction on the size of the body. It should, however, be noted that
at this stage, $\ItZ$ is an arbitrary worldline. Although the momentum $\ItP_\zeta$
evolves along $\ItZ$ by the same formula as  the momentum of a point
charge obeying the Lorentz force law, the center of mass worldline of
the body (defined in the next subsection) does not necessarily
coincide with a Lorentz force trajectory.

Returning to the general case without symmetries, we may obtain general evolution equations for  $p_a$ and $S_{ab}$ by substituting \eqref{pToP} into \eqref{PDot} while using \eqref{D2Xi}. This gives
\begin{subequations}
\label{MomentaEvolve}
\begin{align}
  \frac{\rmD p^a}{\rmd s} =& - \frac{1}{2} R^{a}{}_{bcd} \dot{z}^b S^{cd} + F^a
  \label{PEvolve}
  \\
  \frac{\rmD S^{ab}}{\rmd s} =& ~ 2 p^{[a} \dot{z}^{b]} + N^{ab},
  \label{SEvolve}
\end{align}
\end{subequations}
where $F^a$ and $N^{ab} = N^{[ab]}$ are tensors at $z(s)$ obtained by using \eqref{PDot} to express $\rmd \ItP_\xi/\rmd s$ in terms of the data for the generalized Killing field $\xi^a$ at $z(s)$ via
\begin{align}
  \frac{\rmd \ItP_\xi}{\rmd s} = F^a \xi_a + \frac{1}{2} N^{ab} \nabla_a \xi_b.
  \label{PDotFN}
\end{align}
We refer to $F^a$ and $N^{ab}$, respectively, as the net
force and torque acting on the body. Knowledge of $\rmd \ItP_\xi/\rmd
s$ for all generalized Killing fields $\xi^a$
is completely equivalent to knowledge of $F^a$ and $N^{ab}$.

When the force and torque vanish, \eqref{MomentaEvolve} reduces
precisely to the equations derived by Papapetrou to describe the
motion of a spinning (but otherwise ``structureless'') test particle.
The Papapetrou terms in our equations arise from how the data for a
generalized Killing field vary with $z(s)$, and are thus ``universal'' in the sense that they do not depend on the details of the definition of $\ItP_\xi$.  From \eqref{PDot}, it is clear that both the force and torque vanish entirely in maximally symmetric spacetimes with $F_{ab} = f_a =0$. In this case, the Papapetrou equations are exact for arbitrary extended test bodies. More generally, approximations for the force and torque will
be given in Sect. \ref{Sect: Forces} below. In a multipole series, the lowest order expressions for the gravitational force or torque both involve the quadrupole moment of the body's stress-energy tensor. By contrast, the electromagnetic torque includes dipole and higher moments of $J^a$. The electromagnetic force, of course, starts at monopole (Lorentz) order.

\subsection{Center of mass and motion}
\label{Sect: CM}

As it stands, the formalism just described leaves $\ItZ$ and $n^a(s)$
arbitrary. We now fix these objects in such a way that $\ItZ$ can be interpreted as the body's center of mass worldline.

Under mild assumptions, it may be shown that there exists a unique future-directed unit timelike vector $n_\rmCM^a(z)$ associated with every point $z$ such that
\begin{equation}
  p^{a}(z, n_\rmCM(z)) = m_\rmCM(z) n_\rmCM^a(z)
  \label{nDef}
\end{equation}
for some ``mass'' $m_\rmCM(z) >0$ \cite{Schattner}. Furthermore, there exists a unique timelike
worldline formed out of all points $z_\rmCM$ satisfying
\begin{equation}
  p^a(z_\rmCM , n_\rmCM(z_\rmCM)) S_{ab} (z_\rmCM , n_\rmCM(z_\rmCM)) = 0.
  \label{ZeroDipole}
\end{equation}
The resulting worldline and vector field define what we shall call the
center of mass frame for the given body\footnote{An alternative
  possibility that may be found in the literature sets $n^a
  \propto \dot{z}^a$ and $\dot{z}^a S_{ab} = 0$. This is not, however,
  an acceptable choice. In general, the resulting ``center'' of even a free mass in flat spacetime would be found to satisfy an equation of motion that is third order in time. Its solutions include unphysical helical trajectories as well as the expected
  straight-line motions \cite{Weys}}.
Roughly speaking, \eqref{nDef} defines the preferred hypersurfaces of ``constant time'' to be those attached to an instantaneously zero-momentum ``rest frame.'' Similarly, \eqref{ZeroDipole} essentially implies that the body's mass dipole moment about $z_\rmCM$ vanishes in this same frame. Unless otherwise stated, the given definitions for the reference worldline and foliation will be assumed throughout the remainder of the paper. ``CM'' subscripts will be dropped for brevity when no confusion can arise.

An equation of motion for the center of mass worldline $z(s)$ may be
obtained by differentiating \eqref{ZeroDipole} with respect to $s$ and using methods introduced in \cite{Ehl77}. In doing so, it is convenient to parameterize $\ItZ$ such that $n^a \dot{z}_a = -1$. Note that
this implies that $s$ is {\em not} a proper time along $\ItZ$ unless $n^a$ is
tangent to $\ItZ$. The result is that
\begin{align}
  & m \dot{z}^a = p^a - N^{a}{}_{b} n^b \nonumber
  \\
  & ~ - \frac{S^{ab} [ m \tilde{F}_b + (p^c - N^{c}{}_{d} n^d )(q F_{bc} - \frac{1}{2} S^{df} R_{bcdf}) ]}{ m^2 - \frac{1}{2} S^{bc} (q F_{bc} - \frac{1}{2} S^{df}R_{bcdf} ) } .
  \label{MomVel}
\end{align}
The $m$ appearing here was defined by \eqref{nDef}, so $m = \sqrt{-p_a p^a}$.
We have also introduced a ``reduced force'' $\tilde{F}^a \equiv F^a
- q F^{a}{}_{b} \dot{z}^b$.

The fact that the momentum $p^a$ fails, in general, to equal $m \dot{z}^a$ is fundamental to the results of this paper. When $p^a \neq m \dot{z}^a$, the body may be said to possess ``hidden momentum.'' This concept has arisen in various contexts before (see, e.g., \cite{DeGroot, HiddenMomentum, Jackson, griffiths, ColemanVanvleck}), although it does not appear to have previously been discussed with this generality. Note that the existence of hidden momentum can introduce some confusion regarding what is meant by the ``force'' on a body.  As is clear from \eqref{PEvolve}, we associate the force on a body with its rate-of-change of momentum.  This notion of force is used consistently throughout this paper, but the reader is warned that it is common in the literature to instead label anything contributing to an acceleration as a force.

It is useful to define a spin angular momentum vector $S^a$ associated with the center of mass frame. This is given by
\begin{equation}
    S^a = -\frac{1}{2} \epsilon^{a}{}_{bcd} (p^b/m) S^{cd} .
    \label{SpinDef}
\end{equation}
Since $p^a S_{ab}=0$, $S_{ab}$ can be recovered from $S^a$ via
\begin{equation}
  S_{ab} = \epsilon_{abcd} (p^c/m) S^d .
\end{equation}
The evolution equation \eqref{SEvolve} now reduces to
\begin{equation}
  \frac{\rmD S^a}{\rmd s} = (p^a/m) (\dot{p}_b/m) S^b - \frac{1}{2} \epsilon^{a}{}_{bcd} (p^b/m) N^{cd}  .
  \label{SpinEvolve}
\end{equation}
Eq. \eqref{PEvolve} may be used to replace the $\dot{p}^b$ appearing here. The first term on the right-hand side is responsible for Thomas precession. It arises from the requirement that $p^a S_a = 0$ remain preserved under time evolution.

It should also be noted that the (rest) mass $m$ does not necessarily remain constant under evolution. Using \eqref{MomentaEvolve} and \eqref{nDef}, it is straightforward to show that \cite{Dix70a}
\begin{equation}
  \frac{\rmd m}{\rmd s}  = - \dot{z}^a F_a + (p^a/m) (\dot{p}^b/m) N_{ab}  .
\end{equation}
The right-hand side of this equation can be naturally transformed into a sum of total time derivatives and ``induction terms'' \cite{Dix70a, Dix79}.

We may now summarize how to find the motion of an extended body. Initial data must
first be specified, consisting of the body's center of mass position
$z(\bar{s})$, linear momentum $p^a(\bar{s})$, and angular momentum
$S^{ab}(\bar{s})$---or, equivalently, $S^a(\bar{s})$---at a chosen time
$\bar{s}$. The evolution of these quantities is then determined by the coupled
first order system \eqref{MomentaEvolve} and \eqref{MomVel}. Of
course, these equations are of no use unless one has some knowledge of
$F^a$ and $N^{ab}$. At least in principle, the force and torque may be extracted from \eqref{PDot} and \eqref{PDotFN}. As a practical matter, however, it is normally more useful to expand the force and torque using multipole approximations. The results of such a procedure are given through quadrupole order in \eqref{GravMultipole} and \eqref{EMMultipole} below.

\subsubsection*{Lab frame centroid}

The problem of defining a center of mass worldline may be thought of as the specification of some notion of mass dipole moment that vanishes when evaluated about an appropriate origin. Our choices \eqref{nDef} and \eqref{ZeroDipole} accomplish this without making use of any special structures external to the body itself. The resulting worldline is known to possess a number of useful properties expected of something declared a ``center of mass'' \cite{Dix70a, Ehl77, Dix79, Schattner, SchattStreub}.

Various other possibilities exist if one allows the introduction of structures external to the body itself. Consider, for example, a spacetime endowed with a ``preferred'' timelike vector field $\tau^a$. This might be Killing, but more generally, it may be interpreted as defining a congruence of ``lab frame observers.'' These observers have a 4-velocity field
\begin{equation}
n^a_\rmL = \tau^a / \sqrt{-\tau_b \tau^b} .
\label{nlab}
\end{equation}
The ``constant time'' hypersurfaces $\Sigma(s)$ used to define a body's momenta may now be generated by setting $n^a = n^a_\rmL$. Similarly, \eqref{ZeroDipole} may be replaced by
\begin{equation}
 n^a_\rmL(z_\rmL)  S_{ab}(z_\rmL,n_\rmL( z_\rmL ) ) = 0.
  \label{ZeroDipoleLab}
\end{equation}
The worldline formed by the set of all points $z_\rmL$ satisfying this equation will be referred to as the body's ``lab frame centroid.'' It is interpreted as the set of points about which the body's mass dipole moment vanishes according to the lab frame observers.

It is instructive to compare the lab frame centroid of a given body with the actual center of mass defined by \eqref{nDef} and \eqref{ZeroDipole} above. In order to make this comparison, it will also be helpful to introduce a third, ``hybrid'' definition for a body's center, wherein we take $n^a= n^a_\rmL$ as above, but now consider points $z_\rmH$ satisfying
\begin{equation}
  p^a(z_\rmH, n_\rmL(z_\rmH)) S_{ab}(z_\rmH,n_\rmL(z_\rmH)) = 0.
  \label{ZeroDipoleHybrid}
\end{equation}
Momentum-velocity relations analogous to \eqref{MomVel} are easily
derived for both the lab frame and hybrid centroids. In the hybrid case,
there is no change other than placing subscripts ``$\rmH$'' on all
quantities appearing in that equation. This means that \eqref{MomVel} holds with the substitutions $p_\rmCM^a \rightarrow p^a_\rmH \equiv p^a(z_\rmH, n_\rmL)$, $F^a_\rmCM \rightarrow F^a_\rmH \equiv F^a(z_\rmH, n_\rmL)$, etc. Note that these quantities can differ from their counterparts associated with the center of mass frame due to the different choice of hypersurface and origin.

The momentum-velocity relation in the lab frame case may be obtained by directly differentiating \eqref{ZeroDipoleLab}. This yields
\begin{equation}
    (-\tau_b p^b_\rmL) \dot{z}_\rmL^a = (-\tau_b \dot{z}_\rmL^b ) p^a_\rmL - N^{ab}_\rmL \tau_b - S^{ab}_\rmL \dot{z}^c_\rmL \nabla_c \tau_b,
    \label{centroid}
\end{equation}
where $p^a_\rmL \equiv p^a(z_\rmL,n_\rmL)$, etc. The apparent simplicity of this equation in comparison with \eqref{MomVel} is due to the fact that it has not been manipulated into an explicit formula for $\dot{z}^a_\rmL$.

One may also define spin vectors associated with both the lab and hybrid frames.
$S_\rmH^a$ is naturally given by \eqref{SpinDef} with the usual replacements $p^a_\rmCM \rightarrow p^a_\rmH$, etc. The form of the evolution equation \eqref{SpinEvolve} is clearly preserved in this case. $S^a_\rmL$ is, however, defined slightly differently. Let
\begin{equation}
  S^a_\rmL = - \frac{1}{2} \epsilon^{a}{}_{bcd} n_\rmL^b S^{cd}_\rmL .
\end{equation}
We then have
\begin{equation}
  \frac{\rmD S^a_\rmL}{\rmd s} = n^a_\rmL S_{\rmL,b} \dot{z}^c_\rmL \nabla_c n^b_\rmL - \frac{1}{2} \epsilon^{a}{}_{bcd} n_\rmL^b (N^{cd}_\rmL + 2 p^c_\rmL \dot{z}^d_\rmL).
  \label{SpinEvolveLab}
\end{equation}

The centroid equations \eqref{centroid} and \eqref{SpinEvolveLab} differ from their respective center of mass equivalents \eqref{MomVel} and \eqref{SpinEvolve} mainly in the terms explicitly involving spin. At slow speeds (so $\dot{z}^a_\rmCM \simeq n^a_\rmL$), \eqref{MomVel} reduces to
\begin{equation}
  m_\rmCM \dot{\vec{z}}_\rmCM - \vec{p}_\rmCM =  (\vec{S}_\rmCM \times \vec{F}_\rmCM)/m_\rmCM + \ldots
  \label{MomeVelSpinCM}
\end{equation}
where ``$\dots$'' includes terms nonlinear in the spin and terms independent of the spin (which are not necessarily negligible).
The vector notation used in \eqref{MomeVelSpinCM} denotes a decomposition with respect to an orthonormal triad orthogonal to $\tau^a$. The lab frame equivalent of this equation is obtained from \eqref{centroid}, and reduces to
\begin{equation}
  m_\rmL \dot{\vec{z}}_\rmL - \vec{p}_\rmL = - \vec{S}_\rmL \times \vec{a}_{\mathrm g} + \ldots ,
  \label{MomVelSpinLab}
\end{equation}
where
\begin{equation}
  a_{\mathrm g}^b \equiv - n^a_\rmL \nabla_a n^b_\rmL
  \label{GravAccel}
\end{equation}
is minus the acceleration of a lab frame observer. If $\tau^a$ were Killing, this would correspond to the ``acceleration due to gravity'' in a Newtonian limit.

Eqs. \eqref{MomeVelSpinCM} and \eqref{MomVelSpinLab} display a curious reciprocity.  For a body moving in flat spacetime subject to an external force, there is a spin-dependent contribution to the hidden momentum in the center of mass frame, while there is negligible spin-dependent hidden momentum in the lab frame (since, choosing $\tau^a$ Killing, the ``acceleration due to gravity'' $a^b_{\mathrm{g}}$ is zero).  On the other hand, for a body freely falling in a stationary, curved spacetime (i.e., a body moving under the influence of gravity), the situation is reversed: there is negligible spin-dependent hidden momentum in the center of mass frame (since the force $F^a$ approximately vanishes), but substantial spin-dependent hidden momentum in the lab frame.  When gravity is treated as an external force, it contributes a spin-dependent hidden momentum of the correct form, but it does so in the ``wrong'' frame.

The situation for spin evolution is in fact precisely analogous. At slow speeds, the timelike components of \eqref{SpinEvolve} and \eqref{SpinEvolveLab} may be written as
\begin{equation}
  n_{\rmL,a} \frac{\rmD S^a_\rmCM }{\rmd s} = - (\vec{S}_\rmCM \cdot \vec{F}_\rmCM)/m_\rmCM + \ldots
  \label{SDotTau}
\end{equation}
and
\begin{equation}
  n_{\rmL , a} \frac{\rmD S^a_\rmL }{\rmd s} = \vec{S}_\rmL \cdot \vec{a}_{\mathrm{g}} .
\label{SDotTauLab}
\end{equation}
Again, only terms linear in the spin have been retained.  The term on the right-hand-side of eq. \eqref{SDotTau} gives rise to the well-known Thomas precession.  We see that a similar reciprocity occurs with eqs. \eqref{SDotTau} and \eqref{SDotTauLab} as occurs with \eqref{MomeVelSpinCM} and \eqref{MomVelSpinLab}.  In particular, gravity treated as an external force contributes a Thomas precession term, but does so in the ``wrong'' frame.

Having compared the center of mass and lab frames at the level of time-derivatives of position and spin, we now directly analyze the (undifferentiated) positions $z_\rmCM$ and $z_\rmL$ as well as spins $S^a_\rmCM$ and $S^a_\rmL$. To do so, it is most convenient to first compare these quantities to their counterparts in the hybrid frame.  We restrict attention to the case of a body moving in flat spacetime, so that the two different worldlines and associated quantities may be directly compared.  For simplicity, we consider the case of no electromagnetic interactions, although the general external force density $f_a$ may be nonzero.

We first compute the difference between $z_\rmH$ and $z_\rmL$ on a $t = \mathrm{const.}$ hypersurface in global inertial coordinates $x^\mu = (t, x^i)$. We also choose the lab frame observers such that $\tau^a = \partial/\partial t$. Now, a spatial shift in the origin clearly leaves the linear momentum unchanged. We therefore have $P^\mu \equiv p^\mu_\rmL = p^\mu_\rmH$. It is useful to decompose this momentum such that
\begin{equation}
  P^\mu = \gamma M (n^\mu_\rmL + V^\mu),
\end{equation}
where $V^\mu$ is orthogonal to $n^\mu_\rmL$ and $\gamma = (1-V_i
V^i)^{-1/2}$. Using \eqref{FlatS}---which is equivalent to
\eqref{PDef} and \eqref{pToP} in this case---it is straightforward to
see that
\begin{equation}
  S^{\mu\nu}_\rmL = S^{\mu\nu}_\rmH + 2 (z_\rmH-z_\rmL)^{[\mu} P^{\nu]} .
  \label{SOrigins}
\end{equation}
Contracting both sides with $\tau_{\nu}$ while using \eqref{ZeroDipoleLab} and \eqref{ZeroDipoleHybrid}, this reduces to
\begin{equation}
  M (\vec{z}_\rmL-\vec{z}_\rmH) = \vec{V} \times \vec{S}_\rmH .
  \label{DeltaZ}
\end{equation}
Note that no slow-motion approximations have been made in this equation. Closely related results may be found in textbooks \cite{MTW,DixonSR} (in less general contexts). An equation essentially identical to \eqref{DeltaZ} played a role in the post-Newtonian analysis of Landau-Lifshitz momentum flow during bobbing found in \cite{ThorneBobbing}.

The difference between $S_\rmL^\mu$ and $S_\rmH^\mu$ is also straightforward to compute. Using \eqref{SOrigins}, we find that
\begin{equation}
  S^\mu_\rmH = \gamma ( S^\mu_\rmL + \tau^\mu V_i S^i_\rmL) .
\label{SHL}
\end{equation}
The correction term here gives rise to the Thomas precession of $S^a_\rmH$.
As previously noted, $S^\mu_\rmL$ does not experience Thomas precession in flat spacetime.

We now compare the true center of mass $z_\rmCM$ with $z_\rmH$.  It is clear
that $z_\rmCM = z_\rmH$ for a free mass (i.e., one with external force density $f_a = 0$) in flat spacetime, since the momentum is then independent of the choice of hypersurface. Somewhat more generally, suppose that $f_a \neq 0$, but
that the motion is nonrelativistic in the lab frame, i.e., $|\vec{V}|^2 = V_i V^i \ll 1$. We further assume that $\tau_a f^a \lesssim O(|\vec{V}| | \vec{f}
|)$, as will be the case if the net external force $F^a$ is (nearly)
orthogonal to the $4$-momentum. Finally, suppose that the body's
characteristic size $R$ satisfies
\begin{equation}
| \vec{f} | R^4/M \lesssim O(|\vec{V}|) \ll 1 .
\label{FrM}
\end{equation}
This implies, roughly speaking, that the velocity does not change significantly during a light travel time across the body. It may then be shown that
\begin{equation}
    p^\mu_\rmCM = P^\mu + O(|\vec{V}|^2),
    \label{DeltaP}
\end{equation}
and
\begin{equation}
    S^{\mu}_\rmCM = S^{\mu}_\rmH + O(|\vec{V}|^2).
\label{SHCM}
\end{equation}
It follows immediately that
\begin{equation}
    M \vec{z}_\rmCM = M \vec{z}_\rmH + O(|\vec{V}|^2)
    \label{DeltaZ3}
\end{equation}
Combining \eqref{SHL} and \eqref{SHCM}, we obtain
\begin{equation}
    S^{\mu}_\rmCM  = S^\mu_\rmL + \tau^\mu V_i S^i_\rmL + O(|\vec{V}|^2) .
\end{equation}
This accounts for how $S^{\mu}_\rmCM$ can experience Thomas precession while $S^\mu_\rmL$ does not. Combining \eqref{DeltaZ} and \eqref{DeltaZ3}, we obtain
\begin{equation}
    M \vec{z}_\rmCM  = M \vec{z}_\rmL + \vec{S}_\rmL  \times \vec{V} + O(|\vec{V}|^2)
    \label{DeltaZ2}
\end{equation}
when both worldlines are compared on a single $t = \mathrm{const.}$ hypersurface. This accounts for how there can be hidden momentum associated with spin in the center of mass frame while there is none in the lab frame.
Although our derivation has been restricted to Minkowski spacetime, a comparison of \eqref{MomeVelSpinCM} with \eqref{MomVelSpinLab} suggests that a similar result also holds at least in stationary curved spacetimes (where $\tau^a$ may be chosen to be Killing, so $a^b_\mathrm{g}$ has the interpretation of a ``gravitational acceleration'').

\subsection{Evaluation of forces and torques}
\label{Sect: Forces}

Returning to the general development where spacetime may be curved and $F_{ab} \neq 0$, recall that the force and torque appearing in \eqref{MomentaEvolve} can be extracted from \eqref{PDot} and \eqref{PDotFN} using explicit expressions for the generalized Killing fields in terms of their Killing data. This results in integral expressions that may be found in Eqs. (12.10) and
(12.11) of \cite{Dix74} (in the case $f_a=0$).  These integrals can be
expanded in a multipole series, and we now present the results of this
up to quadrupole order. The formulas in this subsection are valid for
an arbitrary choice of worldline $\ItZ$ and vector field
$n^a$. However, one would not expect the
quadrupole approximation to be valid unless $\ItZ$ is close to the
center of mass worldline and $n^a$ close to the choice \eqref{nDef}. The
metric and electromagnetic field must also vary slowly over each
cross-section of the body. Note that this last assumption implicitly
implies that self-fields are small, i.e., that the object under consideration is a test
body\footnote{As previously mentioned at the beginning of this
  section, modifying \eqref{PDef} and \eqref{Quadrupole} in a
  particular way allows large self-fields to be included very
  naturally \cite{HarteScalar, HarteEM,
    HarteHighMoments}. In this case, the laws of motion presented here
  still hold if only the ``regular'' Detweiler-Whiting component of
  the self-field is sufficiently small. This is normally much smaller
  than the full self-field.}. A more precise discussion of the conditions necessary to ensure an accurate multipole expansion may be found in \cite{Dix67}.

Up to quadrupole order, the gravitational force and torque are \cite{Dix70a, Dix74}
\begin{subequations}
\label{GravMultipole}
\begin{align}
  F^a_{\mathrm{grav}} = - \frac{1}{6} J^{bcdf} \nabla^a R_{bcdf}
  \label{ForceGrav}
  \\
  N^{ab}_{\mathrm{grav}} = - \frac{4}{3} J^{cdf[a} R_{cdf}{}^{b]} ,
  \label{TorqueGrav}
\end{align}
\end{subequations}
where $J^{abcd}$ is the quadrupole moment of $T^{ab}$, defined by \cite{Dix74, HarteHighMoments}
\begin{align}
  J^{abcd} = \int_\Sigma \rmd S_{c'} T^{a'b'} \sigma^{[a} \sigma^{|[c|} \bigg[ \sigma^{b]}{}_{a'} \sigma^{d]}{}_{b'} w^{c'} \nonumber
  \\
  ~ + 2 \Theta^{b]d]fh} H_{a'f} \dot{z}_h \delta^{c'}{}_{b'} \bigg].
  \label{Quadrupole}
\end{align}
The notation here indicates independent antisymmetrizations of the
index pairs $(a,b)$ and $(c,d)$. As
noted before, $\sigma_a = \nabla_a \sigma(x',z)$ denotes
the derivative of Synge's world function and $w^{c'}(x')$ is the time
evolution vector field for the hypersurfaces $\Sigma(s)$. Similarly,
$\sigma_{aa'} = \nabla_a \nabla_{a'} \sigma(x',z)$
and
\begin{equation}
  H^{a'}{}_{a} = [- \sigma^{a}{}_{a'}]^{-1} ,
\end{equation}
where the ``$-1$'' denotes the inverse. We assume that the body is
sufficiently small that this inverse exists on each cross-section $\Sigma \cap W$. Finally, $\Theta^{abcd} (x',z)$ is defined by
\begin{equation}
  \Theta^{abcd} (x',z) = \int_0^1 \rmd u \, (\sigma^{a a''} \sigma^{(c}{}_{a''} \sigma^{d)}_{b''} \sigma^{b b''})|_{(y''(u),z)},
\end{equation}
where $y''(u)$ is an affinely-parameterized geodesic satisfying $y''(0)=z$ and $y''(1) =x'$. In flat spacetime,
\begin{equation}
  \Theta^{abcd} = g^{a(c} g^{d)b} .
\end{equation}
Note also that $J^{abcd} = J^{[ab] cd} = J^{ab [cd]}$ and $J^{[abc]d}=0$.

Following \cite{Ehl77}, $J^{abcd}$ may be decomposed into three simpler components via
\begin{equation}
  J^{abcd} = \ItJ^{abcd} + n^{[a} \ItJ^{b]cd} + n^{[c} \ItJ^{d]ab} - 3 n^{[a} \ItJ^{b][c} n^{d]}.
  \label{StressQuad}
\end{equation}
The tensors introduced here satisfy $\ItJ^{ab} = \ItJ^{(ab)}$, $\ItJ^{abc} = \ItJ^{a[bc]}$, $\ItJ^{abcd} = \ItJ^{[ab]cd} = \ItJ^{ab[cd]}$, and
\begin{align}
  n_b \ItJ^{ab} = n_b \ItJ^{abc} = n_b \ItJ^{abcd} = n_d \ItJ^{abcd} = 0,
  \\
  \ItJ^{[abc]} = \ItJ^{[abc]d}= 0.
\end{align}
$\ItJ^{ab}$, $\ItJ^{abc}$, and $\ItJ^{abcd}$ are interpreted as the mass, momentum, and stress quadrupoles, respectively.

Note that if the spacetime metric is Ricci-flat, the addition of appropriately symmetrized multiples of the metric to the full quadrupole moment can be used to show that the laws of motion are unchanged if $J^{abcd}$ is replaced with an ``effective quadrupole moment'' of the form
\begin{equation}
  J^{abcd}_{\mathrm{TF}} = n^{[a} \ItJ^{b]cd}_{\mathrm{TF}} + n^{[c} \ItJ^{d]ab}_{\mathrm{TF}} - 3 n^{[a} \ItJ^{b][c}_{\mathrm{TF}} n^{d]}.
  \label{StressQuadTF}
\end{equation}
$\ItJ^{ab}_{\mathrm{TF}}$ and $\ItJ^{abc}_{\mathrm{TF}}$ have all of the same index symmetries and orthogonality properties as their non-subscripted counterparts, but are additionally trace-free.

Up to quadrupole order, the electromagnetic force and torque are given
by \cite{Dix74}
\begin{subequations}
\label{EMMultipole}
\begin{align}
  F^a_{\mathrm{EM}} = q F^{a}{}_{b} \dot{z}^b - \frac{1}{2} Q^{bc} \nabla^a F_{bc} + \frac{1}{3} Q^{bcd} \nabla^a \nabla_b F_{cd}
  \label{ForceEM}
  \\
  N^{ab}_{\mathrm{EM}} = 2 Q^{c[a} F^{b]}{}_{c} - \frac{4}{3} Q^{cd[a} ( 2 \nabla_c F^{b]}{}_{d} - \nabla_d F^{b]}{}_{c} ).
  \label{TorqueEM}
\end{align}
\end{subequations}
$Q^{ab} = Q^{[ab]}$ is the body's electromagnetic dipole
moment tensor, and $Q^{abc} = Q^{a[bc]}$ its electromagnetic
quadrupole moment tensor. Definitions for these quantities may be found in Eqs. (8.11)-(8.13) of \cite{Dix70b}. They satisfy
\begin{equation}
  n_a Q^{abc} = Q^{[abc]} = 0.
\end{equation}

The tensors $Q^{ab}$ and $Q^{abc}$ may be decomposed as follows. Let
\begin{equation}
  Q^{ab} = 2 n^{[a} d^{b]} - \epsilon^{abcd} n_c \mu_d,
  \label{DipoleDef}
\end{equation}
where $n_a d^a = n_a \mu^a =0$. We refer to $d^a$ and $\mu^a$ as the electric and magnetic dipole moments, respectively. We also write
\begin{equation}
  Q^{abc} = \frac{1}{2} ( 3 \ItQ^{a[b} n^{c]} - \ItQ^{abc} ),
\end{equation}
where $\ItQ^{ab} = \ItQ^{(ab)}$, $\ItQ^{abc} = \ItQ^{a[bc]}$, and $n_b
\ItQ^{ab} = n_b \ItQ^{abc} = \ItQ^{[abc]} =0$. Note that only the trace-free part of $\ItQ^{ab}$ contributes to the force and torque if $\nabla^a F_{ab}=0$.

In summary, the center of mass worldline of an extended body may be found by solving Eqs. \eqref{MomentaEvolve} and \eqref{MomVel}. Adopting the assumptions outlined above, the gravitational and electromagnetic forces and torques appearing in these equations should be well-described by \eqref{GravMultipole} and \eqref{EMMultipole}, respectively. Note that this system is not complete in that we do not have evolution equations for $Q^{ab}$, $Q^{abc}$, or $J^{abcd}$. This is as expected since such equations should depend on the type of material under consideration. They cannot be deduced from stress-energy and charge-current conservation alone.

As a simple application of these approximations, a formula for the hidden
momentum of an extended charge moving in curved spacetime can be straightforwardly obtained
by substituting \eqref{GravMultipole} and \eqref{EMMultipole} into
\eqref{MomVel}.
Although there is no obstacle
to including additional terms, we shall display the result here only in
the approximation of neglecting all terms of quadrupole and higher order as well as all terms nonlinear in the dipole moments and spin.
One immediately obtains
\begin{equation}
   m \dot{z}_a = p_a + (q \epsilon_{acdf} F_{b}{}^{c} n^d S^f/m - 2 Q^{c}{}_{[a} F_{b]c}) n^b.
   \label{hidmo}
\end{equation}
The content of this equation becomes more clear at slow speeds.  Use of \eqref{DipoleDef} and the usual lab frame decomposition of $F_{ab}$ into electric and magnetic fields (see eq. \eqref{EBDef}) gives
\begin{equation}
  m \dot{\vec{z}} = \vec{p} + \vec{S} \times (q \vec{E}/m) - \vec{\mu} \times \vec{E} - \vec{d} \times \vec{B}.\label{hidmoslo}
\end{equation}
Note that we have already encountered the second term on the right side of this equation in \eqref{MomeVelSpinCM} above.

\section{Simple examples of bobbing motion}
\label{Sect: SimpleBobbing}
We are now in a position to illustrate the presence of bobbing effects in a number of simple systems. First, Sect. \ref{Sect: tether} analyzes the motion of two spinning balls connected by an elastic tether (``tetherballs'') in flat spacetime. Test bodies moving in the electromagnetic and gravitational fields of cylindrically symmetric sources are then considered in Sect. \ref{Sect: CylSym}. For the tetherballs, we will find that the bobbing is of an entirely ``kinematic'' nature.  For the test body moving in the field of a line charge, we will find that there is also a ``dynamical'' contribution to the bobbing.  Finally, for the test body moving in the field of a line mass, we will find that there are important differences in the analysis, but that the final results are very similar to the electromagnetic case.

\subsection{Tetherballs}
\label{Sect: tether}

Consider two solid, uncharged masses A and B (the ``balls'')
connected by a thin elastic band T (the ``tether'') in flat
spacetime. The worldtubes $W_\rmA$ and $W_\rmB$ associated with the two
balls are assumed to be disjoint. We also assume that the thickness of the tether is very small compared to the spatial extent of A or B (in any direction).  The total stress-energy tensor of the system is conserved,
\begin{equation}
  \nabla_b (T^{ab}_\rmA + T^{ab}_\rmB + T^{ab}_\rmT) =0.
\end{equation}
Consequently, if we view A, B, and T as a single system, the
net force and torque vanish. Its center of mass moves at constant velocity in global inertial coordinates and its total linear and angular momenta are both conserved. By \eqref{MomVel}, the center of mass velocity of the complete system is collinear with its linear momentum.

Nevertheless, the center of mass positions $z_\rmA(s)$ and $z_\rmB(s)$
of the \textit{individual} balls---as computed from $T^{ab}_\rmA$ and
$T^{ab}_\rmB$ following Sect. \ref{Sect: CM}---display interesting
behavior. To simplify the problem, suppose that the tether is
connected to A and B at their respective centers of mass\footnote{It
  should be noted that since A and B undergo non-uniform motion, they
  cannot remain ``rigid.'' The material point in the body at which
  the center of mass is located may therefore change with time. Nevertheless,
  for motion that is not particularly relativistic and balls that are not particularly deformable, such effects should be very small. If necessary, we can
  assume that the ball is equipped with a servo-mechanism that
  continually adjusts the location of the tether's attachment to the
  ball to lie at the center of mass.}. It then follows from
\eqref{PDot} and \eqref{PDotFN} that the tether exerts no torque on
either ball:
\begin{equation}
 N^{ab}_\rmA = N^{ab}_\rmB = 0.
\end{equation}
One also sees that the forces on A and B are given by appropriate integrals of $f^a = - \nabla_b T_\rmT^{ab}$. Using \eqref{PEvolve}, \eqref{MomVel}, and \eqref{SpinEvolve}, we find that the laws of motion for each mass are
\begin{align}
  \frac{\rmD p^a}{\rmd s} &= F^a
  \\
  m \dot{z}^a &= p^a - \epsilon^{a}{}_{bcd} F^b p^c S^d/m^2
  \label{MomVelTether}
  \\
  \frac{\rmD S^a}{\rmd s} &= p^a (F^b S_b)/m^2 .
  \label{SpinTether}
\end{align}
Subscripts ``A'' or ``B'' should be appended to each variable here as appropriate.

Despite the relative simplicity of these equations, solving for the general motion of the balls would be extremely difficult in general. The forces $F_\rmA^a$ and $F_\rmB^a$ are unknown, and modeling them would require analyzing the dynamical degrees of freedom of the tether itself.\footnote{In Newtonian mechanics, it would be consistent to
  assume that the tether has negligible mass and is arbitrarily stiff,
  so that it can instantaneously propagate forces from one ball to the
  other. Such assumptions would allow one to ignore the dynamical
  degrees of freedom of the tether. However, in special relativity, if
  the tether satisfies the dominant energy condition, its tension
  must remain less than its mass density. It is would therefore be inconsistent to
  assume a tether of negligible mass exerting arbitrarily large forces. For similar reasons, the tether also cannot propagate forces faster than the speed of light.} In
order to simplify the problem, we proceed as follows. Consider the
global inertial coordinates $(t,x^i)$---which we refer to as the
``lab frame"---for which the center of mass of the total system remains fixed at $x^i=0$. Use the timelike Killing vector $\tau = \partial/\partial t$ to define ``lab frame momenta" $P^\mu = p^\mu_\rmL$ as in Sect. \ref{Sect: CM}. For ball A, we then have
\begin{equation}
 P^\mu_\rmA = \int T^{\mu0}_\rmA \rmd^3 x,
\end{equation}
along with similar expressions for ball B and the tether T. The spatial components of the total lab frame momentum vanish in the chosen frame: $P^i_\rmA + P^i_\rmB + P^i_\rmT = 0$.

We now assume that the balls are identical in their physical properties. We also also make the following additional assumptions:
\begin{enumerate}
    \item{In the lab frame, the motion of the balls and tether is non-relativistic and \eqref{FrM} holds.}
    \item{Each component of the lab frame momenta
      satisfies\footnote{This condition will hold if the mass of the tether is small, and that no portion of it moves much faster than the balls.} $P^\mu_\rmT
      \ll P^\mu_\rmA$.}
    \item{The motion of the system is symmetric with respect to rotations by $\pi$ about the $x^3$-axis.}
\end{enumerate}
Defining $\zeta = \partial/\partial x^3$, our assumptions imply that $P^{\mu}_\rmA + P^{\mu}_\rmB \simeq 0$ and $P^\mu_\rmA \zeta_\mu = P^\mu_\rmB \zeta_\mu$. Hence,
\begin{equation}
  P^\mu_\rmA \zeta_\mu = P^\mu_\rmB \zeta_\mu \simeq 0.
\label{LabFrameP}
\end{equation}
It follows from \eqref{DeltaP} that the vertical momenta associated with the center of mass frame satisfy a similar equation,
\begin{equation}
  p^\mu_\rmA \zeta_\mu = p^\mu_\rmB \zeta_\mu \simeq O(| \vec{V}^2 |).
\label{pA}
\end{equation}

To lowest order in the velocity of the balls, the $x^3$-component of the momentum-velocity relation \eqref{MomVelTether} reduces (for either body, with subscripts $\rmA$ or $\rmB$ dropped) to
\begin{equation}
    p^\mu \zeta_\mu = m \dot{\vec{z}} \cdot \vec{\zeta} - (\vec{S} \times \vec{F}) \cdot \vec{\zeta}/m .
    \label{MomVelTether2}
\end{equation}
Since by \eqref{pA} the left-hand side vanishes (i.e., $\zeta$ momentum is conserved), this equation shows that the center of mass must move in the $x^3$ direction (i.e., ``bob'') to compensate the hidden momentum $-\vec{S} \times \vec{F}/m$.  If both masses are revolving around one another, Eq. \eqref{SpinTether} and the assumption of non-relativistic motion imply that the orbital timescale will always be much shorter than the time over which the spin experiences any significant evolution. Therefore, $\vec{S}$ may be taken as approximately constant for each individual orbit.  In this case it is straightforward to see that the bobbing occurs in phase with the orbit.  The symmetry of the system implies that this effect is identical for both balls, so that there is a net bobbing of the whole orbital plane, just as in numerical simulations of spinning black holes.

In the above analysis the bobbing of tetherballs is seen to be a direct consequence of the existence of the kinematical hidden momentum present in \eqref{MomVelTether2}.  As discussed in section \ref{Sect: CM}, this kinematical momentum may be viewed as arising from the difference between the lab frame centroid and the center of mass.  Along those lines, we can give an alternative, lab-frame-based analysis of the bobbing effect.  The analog of \eqref{MomVelTether2} in the lab frame is simply $\vec{P} \cdot \vec{\zeta} = m \dot{\vec{z}}_\rmL \cdot \vec{\zeta}$ (as follows from \eqref{centroid}), i.e., there is no hidden momentum in the lab frame.  Then by the vanishing of the lab frame $\zeta$ momentum (equation \eqref{LabFrameP}), the orbit of the lab frame centroid is confined to a plane.  However, the center of mass differs from the lab frame centroid by equation \eqref{DeltaZ2}, and therefore bobs in and out of the plane.   This is the approach to understanding the bobbing that was taken in the introduction.  Whether viewed as a consequence of the ``$\vec{S} \times \vec{F}$'' hidden momentum or of the difference between the lab frame centroid and the center of mass, we refer to this sort of bobbing effect as ``kinematical bobbing'' since it does not depend upon the details of the dynamical forces involved.  It will arise in every example we consider.

If we relax the symmetry assumptions and allow the masses to be unequal, otherwise similar assumptions lead to a net bobbing effect,
\begin{equation}
  (m_\rmA \dot{\vec{z}}_\rmA + m_\rmB \dot{\vec{z}}_\rmB) \cdot \vec{\zeta}
  = \left[ \left(\frac{\vec{S}_\rmA}{m_\rmA} - \frac{\vec{S}_\rmB}{m_\rmB} \right) \times \vec{F}_\rmA \right] \cdot \vec{\zeta} .
  \label{TetherAvVel}
\end{equation}
If both masses are comparable, the amplitude of the system's oscillation is of order $| \vec{S} | | \dot{\vec{z}} |/m$ for optimally-aligned spin vectors. This corresponds at least qualitatively to what has been observed in simulations of spinning black holes binaries \cite{ThorneBobbing, RadialInfall, Campanelli07}.

It is particularly interesting to consider the special case where two spinning masses are pulled (almost) directly toward each other. Suppose, for concreteness, that the two bodies A and B are initially placed along
the $x^1$-axis. Also let both of their spins initially be antiparallel
and directed along (or against) the $x^2$-axis. As both masses are
pulled together by the tether connecting them, \eqref{MomVelTether2}
implies that they will develop a net velocity along the $x^3$-axis. If
they experience an inelastic collision, conservation of total momentum
implies that the resulting composite will have zero spatial momentum. It will also have no external forces acting on it, no hidden momentum, and therefore no final velocity. If the forces pulling the two balls together increase with decreasing separation, the transverse velocity of the two balls would be seen to first increase monotonically as they come together, and then reduce very quickly to zero as they collide. This peculiar sequence of events has been observed in numerical simulations of two spinning black holes allowed to fall directly towards one another and coalesce \cite{RadialInfall}.

\subsection{Line sources}
\label{Sect: CylSym}

We now present two further simple examples of bobbing for test bodies moving in static, cylindrically symmetric electromagnetic or gravitational fields.

\subsubsection{Electromagnetism}

Consider the field of an infinite static line charge in flat spacetime. Let $\tau^a$ denote the unit time translation associated with the rest frame of this charge, and $\zeta^a$ the translational Killing field pointing along its axis of symmetry. It is clearly possible to choose a vector potential $A_a = - \varphi \tau_a$ satisfying $\mathcal{L}_\tau A_a = \mathcal{L}_\zeta A_a =0$.

We now analyze the motion of a charged test body placed in this potential.
Since $\zeta^a A_a =0$ and $\nabla_a \zeta^b =0$, it follows from \eqref{ConsLaw} that
\begin{equation}
  \ItP_\zeta = p_a \zeta^a
\label{Pbody}
\end{equation}
is conserved. For simplicity, we restrict attention to the case where $\ItP_\zeta = 0$.

Although $p_a \zeta^a = 0$, the test body will, in general, have
hidden momentum. It can therefore develop a finite center of mass velocity along the axis of symmetry. Neglecting quadrupole and
higher moments while retaining terms linear in $Q^{ab}$ and $S^{ab}$,
we find from \eqref{hidmo} that
\begin{equation}
  m \zeta^a \dot{z}_a = \zeta^{a} n^{b}( q \epsilon_{ab}{}^{cd} F_{cf} S_d  n^f /m  - 2 Q^{c}{}_{[a} F_{b]c}).
\end{equation}
For slow speeds this reduces to (see equation \eqref{hidmoslo})
\begin{equation}
  m \dot{\vec{z}} \cdot \vec{\zeta} = [ \vec{S} \times (q \vec{E}/m) - \vec{\mu} \times \vec{E}] \cdot \vec{\zeta}.
  \label{MomVelLineCharge}
\end{equation}
The term proportional to spin in this equation is of exactly
the same form as the one appearing in \eqref{MomVelTether2}, and gives rise to analogous ``kinematical'' bobbing. In this case, however, there is an additional $\vec{\mu} \times \vec{E}$ hidden momentum, which gives rise to a ``dynamical'' bobbing contribution as well.

It is interesting to compare this dynamical hidden momentum with the $\zeta$-component of momentum carried by the electromagnetic field itself,
\begin{equation}
  \ItP_{\zeta}^{\textrm{EM}} = \int T_{\textrm{EM}}^{03} \rmd^3 x.
\label{PEM}
\end{equation}
The purely electric field of the line charge carries no momentum, while the momentum of the test body's field is (by assumption) negligible.  Therefore, the only contribution to $\ItP_{\zeta}^{\textrm{EM}}$ is from the ``cross-term'' between the line charge and test-body fields.  In the limit of slow motion, this may be computed in the manner used below to obtain \eqref{eq:Px}, yielding
\begin{equation}
  \ItP_{\zeta}^{\textrm{EM}} =  -(\vec{\mu} \times \vec{E}) \cdot \vec{\zeta}.
\label{PEM2}
\end{equation}
Thus, the electromagnetic field momentum is equal and opposite to the dynamical part of the test body's hidden momentum. As the test body orbits the line charge, the electromagnetic field momentum oscillates $180^\circ$ {\it out of phase} with the dynamical part the hidden momentum of the body, and {\it in phase} with the dynamical contribution to its bobbing velocity.

It might appear that there is a failure of conservation of total
$\zeta$-momentum in this case, since the $\zeta$-momentum of the body\footnote{On account of the symmetries in this problem, $\ItP_{\zeta}$ is in fact equal to the ordinary integrated stress-energy momentum $\int T^{03} \rmd^3 x$ to the order of approximation relevant here.} is conserved by itself while the $\zeta$-momentum of the electromagnetic field oscillates.  This is an artifact of our treating the line charge as ``fixed.''  There is, in fact, a Lorentz force density exerted on the line charge by the electric field that results from the motion of the orbiting magnetic dipole, which is straightforwardly integrated to give a total force of $\rmd/\rmd t(\vec{\mu} \times \vec{E}) \cdot \vec{\zeta}$.  If the line charge were of finite length and thickness---so that it would have finite mass and one could meaningfully talk about its momentum---this force would impart to it a momentum $(\vec{\mu} \times \vec{E}) \cdot \vec{\zeta}$ that exactly compensates for \eqref{PEM2}, restoring conservation of total momentum.

The hidden momentum of the line charge may be shown to be negligible in this case, so, in fact, the line charge itself experiences a slight bobbing effect.\footnote{The moving line charge will now produce a magnetic field, so that its self-field may now carry some momentum of its own.  However, we can ensure that this momentum remains negligible by taking a limit as the mass of the line charge becomes very big (at fixed linear charge density).  Note that the mass times velocity of the line charge remains fixed in this limit.} Performing a mass-weighted average of the $\zeta$-velocities of both objects in this system results in an equation analogous to \eqref{TetherAvVel}.  The mass times bobbing velocity of the line charge exactly cancels the dynamical contribution to the mass times bobbing velocity of the orbiting body, leaving only the kinematical ``$\vec{S} \times \vec{a}$'' term.  The analogous conclusion for the case of two orbiting charges will be obtained more explicitly in Sect. \ref{Sect: GeneralEM} below.

\subsubsection{Gravitation}

Very similar effects also arise for an uncharged, spinning test particle moving in the metric of static, cylindrically-symmetric mass described by the Levi-Civita metric
\begin{align}
  \rmd s^2 = -(x^1)^{4 \rho} \rmd t^2 + (x^1)^{4 \rho ( 2 \rho - 1)} [ (\rmd x^1)^2+ (\rmd x^3)^2] \nonumber
  \\
  ~ + \alpha^{-2} (x^1)^{2-4 \rho} (\rmd x^2)^2 .
\end{align}
$x^1$ is the cylindrical radius and $x^2$ an angular coordinate ranging from $0$ to $2 \pi$. The central mass is oriented along the $x^3$-axis. $\rho$ and $\alpha$ are free parameters, the former of which is interpreted as a mass per unit length \cite{LeviCivita}.

Consider the motion of a spinning test body in this metric. The
translational Killing field $\zeta = \partial/\partial x^3$ has nonvanishing gradient, so
it is the quantity
\begin{equation}
  \ItP_\zeta = p_a \zeta^a + \frac{1}{2} S^{ab} \nabla_a \zeta_b
\label{Pgrav}
\end{equation}
that is conserved rather than $p_a \zeta^a$ by itself. Neglecting effects nonlinear in the spin or involving quadrupole or higher multipole moments of the body's stress-energy tensor, it follows from \eqref{hidmo} that $p^a = m \dot{z}^a$. If we consider the case $\ItP_\zeta=0$, the vertical velocity of an orbiting test mass is given by
\begin{equation}
  m \dot{z}^a \zeta_a = - \frac{1}{2} \epsilon^{abcd} \nabla_a \zeta_b n_c S_d .
  \label{LineMassVel}
\end{equation}
A straightforward calculation shows that the only non-vanishing components of $\nabla_a \zeta_b$ are
\begin{equation}
  \nabla_1 \zeta_3 = - \nabla_3 \zeta_1 = 2 \rho (2 \rho-1) (x^1)^{8 \rho^2-4\rho-1} .
  \label{GradZ}
\end{equation}

Combining this equation with \eqref{LineMassVel} determines the vertical velocity of the orbiting body. The result is more suggestive if written in terms of the fictitious gravitational acceleration $a_\mathrm{g}^b$ introduced by \eqref{GravAccel}. Defining this using the timelike Killing field $\partial/\partial t$ and using \eqref{LineMassVel} and \eqref{GradZ}, we obtain
\begin{equation}
  m \dot{z}_a \zeta^a = - (1-2 \rho) \epsilon_{abcd} n^a \zeta^b S^c a^d_\mathrm{g} .
  \label{LineMassVel2}
\end{equation}
The slow speed limit of this equation is
\begin{equation}
  m \dot{\vec{z}} \cdot \vec{\zeta} =  - (1-2 \rho) (\vec{S} \times \vec{a}_\mathrm{g}) \cdot \vec{\zeta},
  \label{LineMassVel3}
\end{equation}
which is similar to \eqref{MomVelTether2} and \eqref{MomVelLineCharge} above. Once again, the bobbing velocity is found to be proportional to a term with the form ``$(\mathrm{spin}) \times (\mathrm{acceleration})$''.

There are, however, several key differences from the previous two cases. First, linear momentum was (nearly) conserved for both the tetherball and electromagnetic examples. This meant that the bobbing could be understood as arising from a time-dependent hidden momentum. In the gravitational case, however, the hidden momentum is negligible. This is, however, compensated by the fact that it is not the linear momentum that is conserved by itself, but rather the particular combination \eqref{Pgrav} of the linear and angular momenta. Second, it should be noted that the notion of acceleration used in
\eqref{LineMassVel2} is not the true acceleration of the body---which
nearly vanishes---but rather the acceleration of a stationary
observer. Finally, we note that $\rho \ll 1$ in the ``Newtonian
limit.'' If $\rho$ is neglected, the gravitational bobbing here can be
seen to have the same absolute value but opposite sign to the kinematical
part of the bobbing found in the tetherball and electromagnetic
examples.

In the next section, we will show that the close analogy between the
electromagnetic and gravitational results is not an artifact of this
particular example. Rules will be derived that allow us to
translate equations for the motion of a test body in a stationary
electromagnetic field into those for a test body in a
weak, stationary gravitational field. We will thereby account for the
sign difference in the $\vec{S} \times \vec{a}_\mathrm{g}$ bobbing of the test
mass in the gravitational case as arising from the {\it same}
kinematical term as in the tetherball and electromagnetic examples
above, together with a dynamical effect that is a factor of two larger and of
opposite sign. As in the electromagnetic case, the \textit{net} bobbing of the
complete system should be given only by the kinematical term. This illustrated directly by Eq. \eqref{eq:Fnetbobbinggrav} below for the case of two compact masses in orbit around one another.

\section{Analogies between electrodynamics and general relativity}
\label{Sect: Analogy}

It has long been known that there are strong similarities between electric and gravitational phenomena. This is clear when comparing Newton's law of gravity to Coulomb's electrostatics. In that case, the only differences between the two theories lie in an overall sign as well as the existence of both positive and negative electric charge. Many aspects of magnetic phenomena are also analogous to gravitational effects found in general relativity.  Although distinctions between Maxwell's and Einstein's theories are extremely important, their many similarities can still be a useful source of intuition.

We now review how---for the case of a stationary, nearly-flat spacetime with sources that are slowly moving and have negligible stresses---the Einstein field equations can be written in a form analogous to Maxwell's equations\footnote{A number of different comparisons can be made between Einstein's and Maxwell's equations. We choose to adopt a restricted post-Minkowski approximation here (see, e.g., \cite{Wald} for a similar treatment of the field equations). Alternatively, equations associated with the post-Newtonian approximation of general relativity may also be successfully compared with equations found in ordinary electrodynamics \cite{ThorneMaxwell}. It is even possible to derive some exact results by comparing various properties of the ``tidal tensors'' $R_{abc}{}^{d}$ and $\nabla_a F_{bc}$ \cite{GEM2} (see also \cite{GEM1,GEMFrames} for additional analogies).}. We shall then consider test body motion and show that the laws of motion governing an extended mass in such a spacetime can be generated by precise rules from the laws of motion describing an object moving in an electromagnetic field in flat spacetime.

\subsection{Field equations}

Consider a spacetime whose metric $g_{ab}$ satisfies Einstein's
equation with stress-energy source $t_{ab}$.
Suppose that $g_{ab}$ is nearly equal to the flat
metric $\eta_{ab}$, so the perturbation $h_{ab} = g_{ab} - \eta_{ab}$
will approximately satisfy the linearized Einstein equation with
source $t_{ab}$. We choose the Lorenz gauge,
\begin{equation}
 \partial^b \bar{h}_{ab} = 0,
\end{equation}
where
\begin{equation}
  \bar{h}_{ab}= h_{ab} - \frac{1}{2} \eta_{ab} \eta^{cd} h_{cd},
\end{equation}
and $\partial_a$ represents the covariant derivative compatible with
$\eta_{ab}$. Unless otherwise noted, we raise and lower indices with
the flat metric. The linearized Einstein equation then
reads \cite{Wald}
\begin{equation}
  \partial^c \partial_c \bar{h}_{ab} = - 16 \pi t_{ab} .
  \label{Boxh}
\end{equation}
The linearized Bianchi identity implies that $t_{ab}$ is conserved with
respect to $\eta_{ab}$, i.e., $\partial^a t_{ab} = 0$.

Let $\tau^a$ be a time translation Killing field of $\eta_{ab}$ with unit norm with respect to that metric. This fixes a ``lab frame.'' We define a mass-energy current $j^a$ with respect to $\tau^a$ by
\begin{equation}
j_a = -t_{ab} \tau^b.
\end{equation}
It follows immediately that $\partial_a j^a = 0$.

We now restrict attention to the case of stationary perturbations so that $\mathcal{L}_\tau \eta_{ab} = \mathcal{L}_\tau h_{ab} = 0$.  We further assume that the components of $t_{ab}$ projected orthogonally to $\tau^a$ are all negligible compared with the components contributing to $j^a$. This means that in the lab frame adapted to $\tau^a$, stresses in the body are all small compared with the local energy and momentum densities.  In global inertial coordinates $(t,x^i)$ where $\tau = \partial/\partial t$, it follows from \eqref{Boxh} and our stationarity assumption that each component of $\bar{h}_{\mu\nu}$ satisfies Laplace's equation with source $-16 \pi t_{\mu\nu}$. Since $t_{ij}$ has been assumed to be negligible, we have $\bar{h}_{ij} \simeq 0$. The trace-reversed metric perturbation therefore has the form
\begin{equation}
  \bar{h}_{ab} = \mathcal{A}_{(a} \eta_{b)c} \tau^c.
  \label{hBar}
\end{equation}
for some potential $\mathcal{A}_a$. As in electromagnetism, $\mathcal{A}_a$ may be used to define a skew-symmetric field strength
\begin{equation}
  \mathcal{F}_{ab} = 2 \partial_{[a} \mathcal{A}_{b]}.
\end{equation}
It immediately follows that
\begin{equation}
  \partial_{[a} \mathcal{F}_{bc]} = 0.
\end{equation}
This is clearly analogous to one of Maxwell's equations.

The other Maxwell equation is nearly, but not exactly, reproduced in linearized
general relativity. Using \eqref{Boxh}, one finds that
\begin{equation}
  \partial^b \mathcal{F}_{ab} = 32 \pi [j_a + \frac{1}{2} \tau_a (\tau^b j_b) ]
\end{equation}
instead of \eqref{Maxwell}. Despite this, the field
strength may still be split up into electric and magnetic components
that do have precise electromagnetic counterparts. In Maxwell's
theory, one normally defines electric and magnetic fields by
\begin{equation}
  F_{ab} = 2 \tau_{[a} E_{b]} + \eta_{abcd} \tau^c B^d
  \label{EBDef}
\end{equation}
and $\tau^a E_a = \tau^a B_a = 0$. We have used $\eta_{abcd}$ here to denote the volume element associated with the flat metric. In the gravitational case, it is useful to define gravitoelectric and gravitomagnetic fields $\mathcal{E}_a$ and $\mathcal{B}_a$ by
\begin{equation}
  \mathcal{F}_{ab} = 8 (\tau_{[a} \mathcal{E}_{b]} + \eta_{abcd} \tau^c \mathcal{B}^d)
\end{equation}
and $\tau^a \mathcal{E}_a = \tau^a \mathcal{B}_a = 0$. Note the
relative factor of two difference between the definitions of
$\mathcal{E}_a$ and $\mathcal{B}_a$ in terms of $\mathcal{F}_{ab}$ as
compared with those for $E_a$ and $B_a$ in terms of $F_{ab}$. Up to
terms nonlinear in the metric perturbation, the equations satisfied by $\mathcal{E}_a$ and $\mathcal{B}_b$ are now identical to the flat-spacetime Maxwell equations for $E_a$ and $B_a$ with source $j^a$.  It should be emphasized that this correspondence holds only for stationary fields. Fundamentally time-dependent phenomena like electromagnetic induction do not appear to have close analogs in general relativity \cite{GEM2}.

\subsection{Comparing laws of motion}
\label{compare}

We now turn our attention to computing the laws of motion satisfied by freely-falling ($F_{ab} = f_a =0$) extended test bodies in weakly-curved spacetimes like those just described. These will be compared with the equations governing test body motion in flat spacetime electromagnetism. The relevant external fields $F_{ab}$ and $\ItF_{ab}$ will both be assumed stationary and source-free near each body's worldtube.

As a result of this analysis, certain substitution rules will be found that transform electromagnetic equations into gravitational ones. It is important to note that the direction of this correspondence cannot be reversed in any natural way. This is due to the fact that the electromagnetic laws of motion depend on all multipole moments of $J^a$ as well as the first two moments of $T^{ab}$ (the mass and spin). By contrast, motion in the gravitational case depends only on the multipole moments associated with the stress-energy tensor. It follows that the constitutive equations (and initial data) required to complete our laws of motion must be fundamentally different in the electromagnetic and gravitational cases. Even at the monopole level, for example, it would be difficult to take a single parameter---the mass---and uniquely transform it into both a mass and charge.

An additional subtlety arises at dipole order. The dipole moment of $T^{ab}$ is essentially $S^{ab}$, while the dipole moment of $J^a$ is $Q^{ab}$. Both of these quantities can affect the motion of a charge in flat spacetime. In general, however, they may have very different structures. The center of mass condition \eqref{ZeroDipole} strongly restricts the form of $S^{ab}$, for example. Physically, there is no need for such a constraint on $Q^{ab}$. Despite this, simple substitution rules can only be found if we do restrict to special cases satisfying $p_a Q^{ab} =0$. This corresponds to considering objects with vanishing electric dipole moment, and is the only case that will be considered here.

We also introduce several approximations and specialize to globally inertial coordinates aligned with the stationary Killing field $\tau = \partial/\partial t$. Assume that the mass $m$, charge $q$, center of mass velocity $\dot{z}^i$, and linear size $R$ of the test body all scale linearly with a small parameter\footnote{Note that
  this does {\it not} correspond to a post-Newtonian (or
  ``post-Coulombic'') expansion. That would take, e.g., $m \sim
  O(\epsilon)$ but $\dot{z}^i \sim O(\epsilon^{1/2})$. See the next
  section.} $\epsilon$. Internal (e.g., rotational) velocities are
allowed to be of order unity, so we have $S_\nu \sim \mu_\nu \sim
O(\epsilon^2)$ and $\ItJ^{\mu\nu}_{\mathrm{TF}} \sim \ItJ^{\mu\nu\lambda}_{\mathrm{TF}} \sim
O(\epsilon^3)$. Time derivatives of dipole and quadrupole
moments will be neglected.

We are interested in calculating the derivative
\begin{equation}
\frac{\rmd}{\rmd s} \equiv \dot{z}^a \partial_a
\label{dds}
\end{equation}
of the body's rest mass $m$, spin $S_\mu$, and ``kinetic momentum'' $m \dot{z}^i$, to order $\epsilon^3$. Only terms linear in the fields $E_a$ and $B_a$ or $\mathcal{E}_a$ and $\mathcal{B}_a$ will be retained. This calculation extends results obtained in \cite{WaldSpin}.  Note that the derivative operator
$\partial_a$ associated with $\eta_{ab}$ appears on the right side of
\eqref{dds}, so in the gravitational case, $\rmd \dot{z}^i/\rmd
s = \rmd^2 z^i/\rmd s^2$ measures the deviation from straight line motion
in the metric $\eta_{ab}$ rather than the deviation from geodesic
motion in the metric $\eta_{ab} + h_{ab}$.

We compute $\rmd^2 z^i/\rmd s^2$ in both the electromagnetic and
gravitational cases by directly differentiating
\eqref{MomVel} while using \eqref{PEvolve} to evolve the momentum. The forces and torques appearing in the resulting equations are those discussed in Sect. \ref{Sect: Forces}. In the gravitational case, these can be simplified in the weak-field, slow motion approximation being considered here. First note that to linear order in the metric perturbation, the Riemann tensor has the form
\begin{align}
  R_{abcd} = \frac{1}{2} ( \tau_{[a} \partial_{b]} \mathcal{F}_{cd} + \tau_{[c} \partial_{d]} \mathcal{F}_{ab}) \nonumber
  \\
  ~ - \partial_{[a} ( \eta_{b][c} \mathcal{F}_{d]f}) \tau^f.
\end{align}
Substituting this into the formula \eqref{ForceGrav} for the gravitational quadrupole force yields
\begin{equation}
    F^a_{\mathrm{grav}} = - \frac{1}{3} J^{bcdf} \partial^{a} \partial_c \ItF_{d [f} \eta_{g] b} \tau^g.
    \label{ForceGravApprox}
\end{equation}
Similarly, the torque \eqref{TorqueGrav} reduces to
\begin{align}
  N^{ab}_{\mathrm{grav}} = \frac{2}{3} J^{cdf [a} [ \tau_c ( 2 \partial_d \ItF^{b]}{}_{f} - \partial_f \ItF^{b]}{}_{d}) \nonumber
  \\
  - (\tau^{b]} \partial_c \ItF_{df} - \eta_{df} \partial_c \mathcal{F}^{b]}{}_{g} \tau^g) ] .
  \label{TorqueGravApprox}
\end{align}
It is useful to replace both instances of $J^{abcd}$ in these formulas
by the ``effective quadrupole moment'' $J^{abcd}_{\mathrm{TF}}$ with
expansion \eqref{StressQuadTF}. We similarly replace the electromagnetic quadrupole components $\ItQ^{ab}$ and $\ItQ^{abc}$ in \eqref{ForceEM} and \eqref{TorqueEM} by their trace-free components $\ItQ^{ab}_{\mathrm{TF}}$ and $\ItQ^{abc}_{\mathrm{TF}}$. For the stationary source-free fields considered here, these substitutions have no effect on the laws of motion.

In the electromagnetic case, we find that the acceleration $\rmd^2
z^i/\rmd s^2$ of a charged body with vanishing electric dipole moment
in Minkowski spacetime is given by
\begin{align}
  m \frac{\rmd^2 z^i}{\rmd s^2}  = q [ ( 1 + \frac{1}{2} | \dot{\vec{z}} |^2 ) \vec{E} + \dot{\vec{z}} \times \vec{B} ]^i
  \nonumber \\
  ~ + \partial^i [ \vec{\mu} \cdot ( \vec{B} - \dot{\vec{z}} \times \vec{E} ) ] + \dot{z}^j \partial_j [
  \vec{E} \times (\vec{\mu} - \frac{q}{m} \vec{S}) ]^i
  \nonumber
  \\
  ~ + \frac{1}{2} \ItQ^{jk}_{\mathrm{TF}} \partial^i \partial_j E_k - \frac{1}{6} \epsilon_{klm} \ItQ^{jkl}_{\mathrm{TF}} \partial^i \partial_j B^m  + O(\epsilon^4).
  \label{EMforce}
\end{align}
This result corrects\footnote{The term $\dot{z}^j \partial_j [ \vec{E}
    \times (\vec{\mu} - \frac{q}{m} \vec{S}) ]^i$ in \eqref{EMforce}
  arises from the ``hidden momentum,'' and was not known to the author
  of \cite{WaldSpin} in 1972. It did not appear in the first edition
  of Jackson's text on electrodynamics, although part of it, namely
  $\dot{z}^j \partial_j ( \vec{E} \times \vec{\mu})^i$, is mentioned
  in later editions \cite{Jackson}. The full term may effectively be
  found in a text by Suttorp and de Groot \cite{DeGroot}, although it
  does not seem to be widely known. In addition, a calculational error
  appears to have been made in Eq. (45) of \cite{WaldSpin}.} Eq. (45) of
\cite{WaldSpin} and extends it to include quadrupole forces and terms
quadratic in $\dot{\vec{z}}$. Similarly, we find
\begin{equation}
  \frac{\rmd}{\rmd s} [ m - (- \vec{\mu} \cdot \vec{B})] = O(\epsilon^4).
\end{equation}
Thus, $m$ is not constant. A constant of motion is obtained to order $\epsilon^3$ if we {\it subtract} from $m$ the usual expression, $-
\vec{\mu} \cdot \vec{B}$, for magnetic dipole energy. See \cite{Dix70a} and
\cite{GrallaHarteWald} for further discussion. Also note that we are following the quite confusing standard convention of using $\partial^i( \vec{\mu} \cdot \vec{B})$ to denote\footnote{The term $\dot{z}^j \partial_j [\vec{E} \times
(\vec{\mu} - q/m \vec{S})]$ is to be interpreted similarly.
However, if we had not neglected the time dependence of
$\vec{\mu}$ and $\vec{S}$, this term would be replaced by
$\dot{z}^\mu \partial_\mu [\vec{E} \times (\vec{\mu} - q/m
\vec{S})]$. The derivative, $\dot{z}^\mu \partial_\mu$, {\it
along} the worldline of $\vec{E} \times (\vec{\mu} - q/m
\vec{S})$ {\it is} well defined, and this derivative would act
on $\vec{\mu}$ and $\vec{S}$ as well as on $\vec{E}$.} $\mu^j \partial^i B_j$. Since $\vec{\mu}$ is defined only on the worldline of the body, it makes no sense to take a spatial derivative of the quantity $\vec{\mu} \cdot \vec{B}$, but there is no good alternative vector notation for $\mu^j \partial^i B_j$.

Very similar calculations may be used to find the coordinate acceleration of a test mass in a stationary, weakly-curved spacetime. This yields
\begin{align}
  m \frac{\rmd^2 z^i}{\rmd s^2} = - m [ (1+ 2 | \dot{\vec{z}} |^2 ) \vec{\mathcal{E}} + 4 \dot{\vec{z}} \times \vec{\mathcal{B}} - 2 (\dot{\vec{z}} \cdot \vec{ \mathcal{E}}) \dot{\vec{z}}  ]^i
  \nonumber
  \\
  ~  - 2 \partial^i [ \vec{S} \cdot ( \vec{\mathcal{B}} - \dot{\vec{z}} \times \vec{\mathcal{E}}) ] - \dot{z}^j \partial_j ( \vec{\mathcal{E}} \times \vec{S})^i
  \nonumber
  \\
  ~ - \frac{1}{2} \mathcal{J}^{jk}_{\mathrm{TF}} \partial^i \partial_j \mathcal{E}_k + \frac{2}{3} \epsilon_{klm} \mathcal{J}^{jkl}_{\mathrm{TF}} \partial^i \partial_j \mathcal{B}^m + O(\epsilon^4),
\label{gravmotion}
\end{align}
which extends Eq. (44) of \cite{WaldSpin} to include quadrupole forces
and terms quadratic in $\dot{\vec{z}}$. In this case, $m$ remains
constant up to $O(\epsilon^3)$.

The evolution of the spin in the electromagnetic and gravitational
cases can be obtained from \eqref{PEvolve}, \eqref{SpinEvolve},  and
the multipole expansions for the force and torque discussed in Sect. \ref{Sect: Forces}. One subtlety in the gravitational case is that we are writing evolution equations in terms of a fictitious flat background metric $\eta_{ab}$. The form of such an equation for $S^\mu$ will therefore be different from one for $S_\mu = g_{\mu\nu} S^\nu$. Of course, the same comment could also be made for the acceleration. There, however, it is clearly most natural to consider $\dot{z}^i$. It is less obvious that $S^\mu$ should be more interesting than $S_\mu$ (or vice versa). We therefore write evolution equations for orthonormal frame components of the spin. In the electromagnetic case, we choose the lab frame basis $\bar{e}_i =
\partial/\partial x^i$, which is orthonormal in the metric $\eta_{\mu
  \nu}$. In the gravitational case, we instead use
\begin{equation}
  e^\mu_i =\bar{e}^\mu_i - \frac{1}{2} \eta^{\mu\nu} h_{\nu\lambda} \bar{e}^\lambda_i  .
\end{equation}
This is orthonormal to linear order in $h_{\mu \nu}$ with respect to the full metric $g_{\mu \nu}$.

A direct calculation now shows that the spin of a charged particle in flat spacetime evolves via
\begin{align}
  \frac{\rmd}{\rmd s} (S_\mu \bar{e}^\mu_i) = [\vec{\mu} \times (\vec{B} - \dot{\vec{z}} \times \vec{E} ) ]_i + \frac{q}{m} ( \vec{E} \cdot \vec{S} ) \dot{\vec{z}}_i \nonumber
  \\
  + \epsilon_{ijl} \ItQ^{jk}_{\mathrm{TF}} \partial_k E^l - \frac{2}{3} \ItQ_{\mathrm{TF}}^{jk}{}_{i} \partial_j B_k + O(\epsilon^4) .
\label{EMspin}
\end{align}
Similarly, the spin of a small mass in a weakly curved spacetime satisfies
\begin{align}
  \frac{\rmd}{\rmd s} (S_\mu e^\mu_i) = - 2 [\vec{S} \times (\vec{\mathcal{B}} - \dot{\vec{z}} \times \vec{\mathcal{E}} )]_i - (\vec{\mathcal{E}} \cdot \vec{S} ) \dot{\vec{z}}_i \nonumber
   \\
   - \epsilon_{ijl} \mathcal{J}^{jk}_{\mathrm{TF}} \partial_k \mathcal{E}^l + \frac{8}{3} \mathcal{J}_{\mathrm{TF}}^{kj}{}_{i} \partial_j \mathcal{B}_k + O(\epsilon^4).
\label{gravspin}
\end{align}
Note that the forms of the Thomas precession terms $q/m(\vec{E} \cdot \vec{S} ) \dot{\vec{z}}_i$ and $-(\vec{\mathcal{E}} \cdot \vec{S})\dot{\vec{z}}_i$ differ from the more familiar ones given in, e.g. \cite{Jackson} and \cite{WaldSpin}. This is because we work here with a tetrad aligned with the lab frame rather than rest frame 4-velocity. Up to this difference, the equations here agree with and extend those of \cite{WaldSpin}.

By inspection, the following recipe allows one to
transform the electromagnetic equations \eqref{EMforce} and
\eqref{EMspin} into the corresponding gravitational ones \eqref{gravmotion} and
\eqref{gravspin}:
\begin{enumerate}
  \item{Identify all terms explicitly involving $S_i$. Replace $q E_i$ by $-m \mathcal{E}_i$, and then leave as-is.}
  \item{To all other terms, send $m \rightarrow m$, $q \rightarrow m$, $\mu^i \rightarrow S^i$, $E_i \rightarrow \mathcal{E}_i$, $B_i \rightarrow 2 \mathcal{B}_i$, $\dot{z}^i \rightarrow 2 \dot{z}^i$, $\ItQ^{ij}_{\mathrm{TF}} \rightarrow \ItJ^{ij}_{\mathrm{TF}}$, $\ItQ^{ijk}_{\mathrm{TF}} \rightarrow 2 \ItJ^{ijk}_{\mathrm{TF}}$. Multiply the overall result by ``$-1$'' (excluding the spin terms set aside in step one).}
  \item{In the expression for acceleration in the gravitational case, add the term $2m \dot{z}^i ( \dot{z}^j \mathcal{E}_j)$.}
\end{enumerate}
The first of these rules essentially states that all spin-dependent terms in the electromagnetic equations are preserved in the gravitational case (with the force $qE_i$ replaced by the analogous force $-m\mathcal{E}_i$). These are exactly the same terms argued to be kinematical in Sect. \ref{Sect: CM}, and the nature of this substitution further supports that interpretation. Our second replacement rule is more familiar. It states that dynamical effects in electromagnetism tend to have the opposite sign in gravity. Velocity-dependent dynamical terms also tend to be twice as important in gravity as they are in electromagnetism (hence $B_i \rightarrow 2 \mathcal{B}_i$, $\dot{z}^i \rightarrow 2 \dot{z}^i$, and $\ItQ^{ijk}_{\mathrm{TF}} \rightarrow 2 \ItJ^{ijk}_{\mathrm{TF}}$).

In addition to this, there is one term in the gravitational acceleration, namely $2m \dot{z}^i ( \dot{z}_j \mathcal{E}^j)$, that has no electromagnetic counterpart. The gravitational and electromagnetic equations cannot be made to agree perfectly at all orders, and the presence of this term signals that we have carried the analogy about as far as it will go (at least in this direction).

Several points of caution should be noted about our substitution rules. First, they provide a recipe \textit{only} for transforming one particular set of differential equations into another. Solutions to the gravitational equations are not necessarily related in any simple way to those of the electromagnetic equations. The given rules also apply only to the center of mass acceleration and spin evolution equations. Momentum-velocity relations, forces, and torques cannot be reliably transformed in the same way. Even the center of mass acceleration and spin evolution equations will not necessarily correspond as we have outlined if a different gauge is chosen for $h_{\mu\nu}$ or a different tetrad is used to decompose $S^\mu$. Finally, the correspondence rules apply only to the equations obtained at the stated order of approximation.

\section{Electromagnetic Binary System}
\label{Sect: GeneralEM}

We now move to our electromagnetic analog of a black hole binary, which consists of two charged magnetic dipoles with spin orbiting around one another in flat spacetime.  Working in a slow-motion regime analogous to the post-Newtonian limit of general relativity, we first analyze the bobbing motion of this system, and then explore the flow of momentum between the field and the bodies.  Finally, we use this information to speculate about kicks.  At each stage in this process, we discuss what conclusions may be drawn for the gravitational case.  As in previous sections, $3$-vector notation will be used to denote the spatial components of $4$-vectors in global inertial coordinates.

\subsection{Laws of motion}

For ease of comparison with the gravitational case, we adopt a slow-motion limit of electrodynamics that is closely analogous to the standard post-Newtonian limit of general relativity (as applied to compact objects).  Introducing a small dimensionless parameter $\lambda$, we scale each body's 3-velocity, mass, charge, spin, and magnetic dipole moment such that
\begin{equation}
    \dot{z}^i \sim O(\sqrt{\lambda}), \, q \sim m \sim O(\lambda), \, S_\nu \sim \mu_\nu \sim O(\lambda^2).
    \label{pcrules}
\end{equation}
Attention will also be restricted to the case of no electric dipole moment, and all quadrupole and higher moments will be assumed $O(\lambda^3)$ or smaller. We also assume that $\rmd \mu_i /\rmd s \sim O(\lambda^3)$ (the same order of magnitude of changes in spin).  Accelerations that contribute to the bobbing effect first appear at $O(\lambda^{2.5})$.  Since this is 1.5 powers of $\lambda$ higher than the Coulomb acceleration, we will call this order of approximation ``1.5 post-Coulombic,'' or 1.5PC, in analogy with terminology used in the post-Newtonian limit of general relativity.

We determine the acceleration of each body by demanding that it move as a test-body in the other body's field, according to the formalism developed in this paper.  Taking into account that the electric field of each body is $O(\lambda)$ and the magnetic field of each body is $O(\lambda^{1.5})$, equations \eqref{PEvolve}, \eqref{ForceEM}, \eqref{DipoleDef}, \eqref{hidmoslo}, and \eqref{EBDef} combine to give at 1.5 post-Coulombic order that
\begin{align}
    m \frac{\rmd^2 z^i}{\rmd s^2} & = q  (1 + \frac{1}{2}  | \dot{\vec{z}}|^2 ) ( \vec{E} + \dot{\vec{z}} \times \vec{B})^i \nonumber \\
    & \quad +  \partial^i [ \vec{\mu} \cdot (\vec{B} - \dot{\vec{z}} \times \vec{E} ) ] \nonumber \\
    & \quad - \frac{\rmd}{\rmd s} [ - \vec{S} \times (q \vec{E}/m) + \vec{\mu} \times \vec{E} ]
    \label{eq:non-rel}
\end{align}
for the acceleration of each body caused by the electric and magnetic fields of the other body.  Note the similarity of this equation to \eqref{EMforce}, which is valid only for stationary external fields.  In fact, equation \eqref{eq:non-rel} could have been derived from \eqref{EMforce} by noting that the field of a body in its instantaneous rest-frame is stationary to the relevant order, using \eqref{EMforce}, and transforming to the lab frame.  Finally, equation \eqref{eq:non-rel} also follows from the work of \cite{HarteEM} and \cite{GrallaHarteWald}, where self-fields were properly taken into account.\footnote{In fact, it can be seen from these references that the Abraham-Lorentz-Dirac self-force contributes to $\rmd^2 z^i/\rmd s^2$ at 1.5PC order. This is unlike the gravitational case, where radiation reaction effects first appear at 2.5PN order.  However, the self-force is always directed along the orbital plane and therefore cannot give bobbing.  We therefore ignore the self-force effects on the orbit.}

The first line of equation \eqref{eq:non-rel} is the ordinary Lorentz force.  The second line is the ordinary dipole force (which too arises from the Lorentz force density).  The third line corrects the integrated Lorentz force density for the presence of hidden momentum of the kinematical (first term) and dynamical (second term) variety.

We now plug in the actual $E$ and $B$ fields to equation \eqref{eq:non-rel} in order to explicitly determine the acceleration.  Formally, we begin with the  Li\'{e}nard-Wiechert field (and its analog for a magnetic dipole) and expand in powers of the velocity.  However, only terms linear in velocity will survive at 1.5PC, so that retardation effects do not enter.  We also specialize here to the terms in the acceleration proportional to dipole and spin.  (These terms are entirely responsible for bobbing, since in the absence of spin and magnetic dipole moment the orbit would be confined to a plane.)  At 1.5PC order, the contributions to the acceleration of body $\rmA$ proportional to a magnetic dipole moment or spin are
\begin{align}
m_\rmA \vec{a}_\rmA^{\mu,S} & = \frac{q_\rmA}{r^3} \left[ 3(\vec{v} \times \hat{r})(\vec{\mu}_\rmB \cdot \hat{r}) - \vec{v} \times \vec{\mu}_\rmB \right] \nonumber \\
& + \frac{q_\rmB}{r^3} \left[ 3((\vec{\mu}_\rmA \times \vec{v}) \cdot \hat{r}) \hat{r} + 3(\hat{r} \cdot \vec{v})(\vec{\mu}_\rmA \times \hat{r}) - 2 \vec{\mu}_\rmA \times \vec{v} \right] \nonumber \\
& + \frac{q_\rmA q_\rmB}{m_\rmA r^3} \left[ (\vec{S}_\rmA \times \vec{v} - 3(\vec{v} \cdot \hat{r})(\vec{S}_\rmA \times \hat{r}) \right],
\label{fdipspin}
\end{align}
where $\vec{r} \equiv \vec{z}_A - \vec{z}_B$ is the orbital separation, and $\hat{r} \equiv \vec{r}/r$. We have also used $\vec{v} \equiv \dot{\vec{z}}_\rmA - \dot{\vec{z}}_\rmB$ to denote the relative velocity of the two charges.  The first line of \eqref{fdipspin} corresponds to the first line of \eqref{eq:non-rel}, i.e., the ordinary Lorentz force on body A due to the dipole field of body B.  The second line of \eqref{fdipspin} corresponds to the second line and final term of the last line of \eqref{eq:non-rel}, i.e., the dipole effects on body A due to the Li\'{e}nard-Wiechert field of body B.  The final line of \eqref{fdipspin} corresponds to the first term of the final line of \eqref{eq:non-rel}, i.e., the kinematical spin effect.

We now focus on bobbing by considering only the acceleration perpendicular to the ``instantaneous orbital plane'' spanned by $\vec{r}$ and $\vec{v}$. Let
\begin{equation}
a_\rmA^{\perp} \equiv \vec{a}_\rmA \cdot \frac{\hat{r} \times \vec{v}}{| \hat{r} \times \vec{v} |}.
\end{equation}
Since only terms proportional to spin and dipole can contribute to $\vec{a}_\rmA^{\perp}$, this quantity may be computed from equation \eqref{fdipspin}.  We find
\begin{align}\label{eq:Fbobbing}
    m_\rmA a_\rmA^{\perp} & = \frac{q_\rmA \vec{\mu}_\rmB + q_\rmB \vec{\mu}_\rmA}{r^3 | \hat{r} \times \vec{v} |} \cdot \left[ \hat{r} ( -2 v^2 + 3 (\hat{r} \cdot \vec{v})^2 ) - \vec{v}(\hat{r}\cdot\vec{v}) \right] \nonumber \\
 & + \frac{q_\rmA q_\rmB \vec{S}_\rmA/m_\rmA}{r^3 | \hat{r} \times \vec{v} |} \cdot [ \hat{r} (v^2 -3(\hat{r} \cdot \vec{v})^2 ) + 2\vec{v}(\hat{r} \cdot \vec{v}) ].
\end{align}
This gives the acceleration out of the plane on each individual body, thus showing that there is a ``dynamical'' contribution (involving $\vec{\mu}_\rmA$ and $\vec{\mu}_\rmB$) in addition to the kinematical effect (involving $\vec{S}_\rmA$).  As in Sect. \ref{Sect: tether}, we now consider the mass-weighted average $m_\rmA a_\rmA^\perp$ + $m_\rmB a_\rmB^\perp$ of the bobbing accelerations, which provides a measure of the net oscillation of the system's orbital plane.  This quantity is easily read off of equation \eqref{eq:Fbobbing} by noting that $\vec{v}$ and $\hat{r}$ are odd under the exchange $\rmA \leftrightarrow \rmB$.  This ``antisymmetrizes'' the kinematical contribution but in fact \textit{eliminates} the dynamical contribution, so that the net bobbing is given by
\begin{align}\label{eq:Fnetbobbing}
    m_\rmA a_\rmA^{\perp} + m_\rmB a_\rmB^{\perp} = \frac{q_\rmA q_\rmB}{r^3 | \hat{r} \times \vec{v} |} \left( \frac{ \vec{S}_\rmA}{m_\rmA} - \frac{\vec{S}_\rmB}{m_\rmB} \right) \nonumber \\ \cdot \left[ \hat{r}\left(v^2-3(\vec{v}\cdot\hat{r})^2\right) + 2\vec{v}(\vec{v}\cdot\hat{r}) \right].
\end{align}
Thus there is in fact \textit{no} net dynamical bobbing in the electromagnetic binary: any dipole effect pushing one body upwards is always compensated by a corresponding dipole effect pushing the other body downwards.  This cancellation is analogous to that found in section \ref{Sect: CylSym}; the dynamical hidden momentum bobbing effect on one body is compensated by the Lorentz force on the other body.  Although this cancelation is somewhat mysterious from the laws of motion point of view, in the next section its necessity will be seen at the level of conservation of momentum.

Recalling that $\vec{E}_\rmB = q_\rmB \hat{r}/r^2 + O(\lambda^{1.5})$, equation \eqref{eq:Fnetbobbing} may be recast back into the canonical ``$\vec{S} \times \vec{F}$'' form of the kinematical bobbing effect (present already in \eqref{eq:non-rel}).  This yields an expression for the net bobbing,
\begin{align}
& m_\rmA a_\rmA^{\perp} + m_\rmB a_\rmB^{\perp} = \nonumber \\
& \frac{\rmd}{\rmd s}
    \Bigg\{ \Bigg[ \left( \frac{ \vec{S}_\rmA}{m_\rmA} - \frac{\vec{S}_\rmB}{m_\rmB} \right) \times (q_\rmA \vec{E}_\rmB) \Bigg] \cdot \frac{ \hat{r} \times \vec{v} }{ | \hat{r} \times \vec{v} | }  \Bigg\},\label{eq:Fnetbobbing2}
\end{align}
which is directly analogous to the net bobbing result obtained in Eq. \eqref{TetherAvVel} for the tetherballs system.

\subsubsection*{Gravity}

The corresponding results for a gravitational binary may be obtained as follows.  As already remarked, in the instantaneous rest frame of (say) body $\rmB$, the fields are sufficiently stationary that the analysis of section \ref{Sect: Analogy} for the motion of body $\rm A$ is valid to 1.5PC and 1.5PN orders.  As in the electromagnetic case, the terms in the acceleration of each body that are linear in velocity can depend only on the relative velocity of the bodies.  Therefore, the substitution rules of Sect. \ref{compare} should hold for equation \eqref{eq:non-rel} restricted to terms linear in the velocity.  However, no non-linearities in velocity occur proportional to spin or dipole, so we can use our substitution rules on the spin and dipole terms of \eqref{eq:non-rel} and plug in the appropriate gravitoelectric and gravitomagnetic fields to obtain the analog of \eqref{eq:Fbobbing},
\begin{align}\label{eq:Fbobbinggrav}
     m_\rmA & a_\rmA^{\perp} = \nonumber \\
 & \frac{-2(m_\rmA \vec{S}_\rmB + m_\rmB \vec{S}_\rmA)}{r^3 | \hat{r} \times \vec{v} |} \cdot \left[ \hat{r} ( -2 v^2 + 3 (\hat{r} \cdot \vec{v})^2 ) - \vec{v}(\hat{r}\cdot\vec{v}) \right] \nonumber \\
 & - \frac{m_\rmB \vec{S}_\rmA}{r^3 | \hat{r} \times \vec{v} |} \cdot [ \hat{r} (v^2 -3(\hat{r} \cdot \vec{v})^2 ) + 2\vec{v}(\hat{r} \cdot \vec{v}) ].
\end{align}
Although all terms are now proportional to spin, we would still say that the first line is dynamical in origin while the second line is kinematical in origin.  The net bobbing again arises only from the kinematical terms,
\begin{align}\label{eq:Fnetbobbinggrav}
   & m_\rmA a_\rmA^{\perp} + m_\rmB a_\rmB^{\perp} = \nonumber \\
 & \frac{m_\rmA \vec{S}_\rmB - m_\rmB \vec{S}_\rmA}{r^3 | \hat{r} \times \vec{v} |} \cdot \left[ \hat{r}\left(v^2-3(\vec{v}\cdot\hat{r})^2\right) + 2\vec{v}(\vec{v}\cdot\hat{r}) \right].
\end{align}
This result agrees with the corresponding equation (5.19) obtained in \cite{ThorneBobbing} using post-Newtonian methods (after dotting the latter equation with $\hat{r} \times \vec{v}/|\hat{r} \times \vec{v}|$ and fixing a typographical error involving a missing factor of $1/r^3$).  Reintroducing the gravitoelectric field $\vec{\mathcal{E}}_\rmB \simeq m_\rmB \hat{r}/r^2$, equation \eqref{eq:Fnetbobbinggrav} may alternatively be written
\begin{align}\label{eq:Fnetbobbinggrav2}
   & m_\rmA a_\rmA^\perp + m_\rmB a_\rmB^\perp = \nonumber \\
& \frac{\rmd}{\rmd s} \Bigg\{ \Bigg[ \left( \frac{\vec{S}_\rmA}{m_\rmA} - \frac{\vec{S}_\rmB}{m_\rmB} \right) \times ( - m_\rmA \vec{\mathcal{E}}_\rmB) \Bigg] \cdot \frac{ \hat{r} \times \vec{v} }{ | \hat{r} \times \vec{v} | } \Bigg\}.
\end{align}
Equations \eqref{TetherAvVel}, \eqref{eq:Fnetbobbing}, and \eqref{eq:Fnetbobbinggrav2} are of an essentially identical form and clearly demonstrate the identical (purely kinematical) nature the net bobbing displayed by slow-motion binaries interacting via elastic, electromagnetic, and gravitational forces.  In each case, the net bobbing is controlled by the identical form of ``$\vec{S} \times \vec{F}$'', where $\vec{F}$ is the particular force holding the bodies in orbit.

\subsection{Momentum flow}
\label{MomFlow}

Although the net bobbing occurring in our electromagnetic binary system is purely kinematical in nature, we now demonstrate that there is a considerable flow of momentum between the bodies and the electromagnetic field during each orbit.

In order to calculate the momentum contained in the electromagnetic field, we note that the stress-energy tensor $T^{ab}_{\mathrm{EM}}$ of the electromagnetic field is quadratic in $F_{ab}$. It naturally breaks up into terms quadratic in the field of body $\rmA$, terms quadratic in the field of body $\rmB$, and cross terms linear in the fields of both bodies $\rmA$ and $\rmB$. By the Schwartz inequality, the self-field contributions to the electromagnetic stress-energy tensor cannot be negligible compared with the cross terms (which we shall denote by $T_\times^{ab}$). Unfortunately, however, inclusion of these self-field terms would take us beyond the test body approximation, and therefore outside the scope of the formalism developed in this paper. We will simply appeal to the work of \cite{HarteEM} and \cite{GrallaHarteWald}, which demonstrates that, at the order of approximation to which we work, stress-energy components quadratic in the field of body $\rmA$ can naturally be viewed as ``self-energy'' terms that comprise part of the stress-energy of body $\rmA$. Similarly, terms quadratic in the field of body $\mathrm{B}$ can be viewed as comprising part of its stress-energy. ``Renormalizing'' both objects' stress-energy tensors in this way, the only self-field effect remaining in the laws of motion at this order is the Abraham-Lorentz-Dirac self-force. This makes no contribution to bobbing or to momentum flow perpendicular to the orbital plane. All self-field effects will therefore be ignored here (as they have been above).

We therefore focus attention on the momentum, $P_\times^i$, carried by the cross-term portion, $T_\times^{ab}$ of the electromagnetic stress-energy tensor,
\begin{equation}
P_\times^i = \int T_\times^{0i} .
\end{equation}
In vector notation, we have
\begin{equation}
\vec{P}_\times = \frac{1}{4 \pi} \int (\vec{E}_\rmA \times \vec{B}_\rmB +  \vec{E}_\rmB \times \vec{B}_\rmA ) \rmd^3 x .
\label{moint}
\end{equation}
To the accuracy required for the 1.5PC approximation (which is $O(\lambda^3)$ for the momentum), we may replace $\vec{E}_\rmA$ by $- \vec{\nabla} \Phi_\rmA$, where $\Phi_\rmA$ is the instantaneous Coulomb potential of body $\rmA$. Making a similar replacement for $\vec{E}_\rmB$ and integrating by parts, we obtain
\begin{equation}
\vec{P}_\times = \int \left( \Phi_\rmA (\vec{J}_\rmB + \frac{\partial \vec{E}_\rmB}{\partial t} ) + \Phi_\rmB ( \vec{J}_\rmA + \frac{\partial \vec{E}_\rmA}{\partial t} ) \right) \rmd^3 x .\label{pint}
\end{equation}
It is now useful to perform the expansion
\begin{equation}
\Phi_\rmA (\vec{x}) =  \Phi_\rmA (\vec{z}_\rmB) - (\vec{x} - \vec{z}_\rmB) \cdot  \vec{E}_\rmA( \vec{z}_\rmB ) + \dots,
\end{equation}
where $\vec{z}_\rmB$ denotes the instantaneous center of mass of body $\rmB$.  Plugging this expansion into the integral of \eqref{pint}, using the definition of the magnetic dipole moment, and keeping only terms proportional to dipole, we obtain
\begin{equation}\label{eq:Px}
\vec{P}^{\textrm{dipole}}_\times = -\vec{\mu}_\rmB \times \vec{E}_\rmA (\vec{z}_\rmB) - \vec{\mu}_\rmA \times \vec{E}_\rmB (\vec{z}_\rmA) .
\end{equation}
Since only momentum proportional to dipole can point out of the orbital plane, this equation (dotted into the normal) gives the entire electromagnetic momentum perpendicular to the plane.  We have thus found that the relevant electromagnetic field momentum is in fact equal and opposite to the sum of the ``$\vec{\mu} \times \vec{E}$'' portions of the hidden momenta associated with the two charges (see equation \eqref{hidmoslo}).

The reason for this remarkable-looking cancellation can be understood as follows. Consider the case where both the 3-current $\vec{J}_\rmA$ of body $\rmA$ and the charge density $\rho_\rmB$ of body $\rmB$ vanish, so, in particular, $\vec{\mu}_\rmA$ and $q_\rmB$ vanish. There will be no force on either body if they are both stationary, so this system can exist in stationary equilibrium. Nevertheless, despite the fact that the bodies and their electromagnetic fields are stationary, the above calculation shows that the cross term in the electromagnetic stress-energy tensor still carries a momentum
\begin{equation}
\vec{P}_\times = -\vec{\mu}_\rmB \times \vec{E}_\rmA (\vec{z}_\rmB)  .
\label{Pfield}
\end{equation}
However, the total stress-energy
\begin{equation}
T_{\textrm{tot}}^{ab} = T_\rmA^{ab} + T_\rmB^{ab} + T_{\rmA,{\rm EM}}^{ab} + T_{\rmB,{\rm EM}}^{ab} + T_{\times}^{ab}
\end{equation}
remains conserved, where $T_\rmA^{ab}$ is the stress-energy of the matter that makes up body $\rmA$ and $T_{\rmA,{\rm EM}}^{ab}$ is the stress-energy of its electromagnetic self-field (and likewise for $\rmB$). In particular, we have
\begin{equation}
\partial_0 T_{\textrm{tot}}^{00} + \partial_i T_{\textrm{tot}}^{i0}  = 0.
\end{equation}
The first term vanishes by stationarity. Multiplying by $\vec{x}$, integrating over space, and then integrating the resulting expression by parts, we obtain
\begin{equation}
P_{\textrm{tot}}^i \equiv \int T_{\textrm{tot}}^{i0} \rmd^3 x = 0 .
\label{Ptot}
\end{equation}
Note that neither the purely electric self-field of body $\rmA$ nor the purely magnetic self-field of body $\rmB$ can make a contribution to $\vec{P}_{\textrm{tot}}$. Consequently, \eqref{Ptot} can be made compatible with \eqref{Pfield} only by assigning a hidden mechanical momentum $\vec{\mu}_\rmB \times \vec{E}_\rmA$ to body $\rmB$
that is equal and opposite to the electromagnetic field momentum. Indeed, this way of resolving the apparent conflict between  \eqref{Ptot} and \eqref{Pfield} was used in \cite{ColemanVanvleck} to argue for the existence of hidden momentum in electromagnetic systems.  Note that in this stationary example, one may construct a simple model of a magnetic dipole to understand at a microscopic level the way in which the hidden mechanical momentum is stored by the dipole \cite{griffiths}.

Given that exact cancellation of the electromagnetic field momentum and hidden mechanical momentum must occur in the stationary case, it should not be surprising that for quasi-stationary systems, there remains a cancellation of the electromagnetic field momentum proportional to the dipole and the non-kinematical portion of the hidden momentum of the bodies.

The above considerations yield a rather surprising picture of what happens in the 1.5PC approximation when two charged, spinning bodies orbit each other with their spins and magnetic dipole moments lying in the orbital plane. During an orbit, there is a substantial transfer of momentum perpendicular to the orbital plane between the electromagnetic field and the bodies. Nevertheless, when the momentum is transferred from the electromagnetic field to the bodies, it does not produce any net bobbing motion; all of the perpendicular momentum transferred to the bodies from the fields goes into the hidden momentum of the bodies, and there is no net effect on the perpendicular motion of their centers of mass.  Net bobbing arises solely from the kinematic ``$\vec{a} \times \vec{S}$'' contribution to the hidden momentum---exactly as for the tetherballs of subsection \ref{Sect: tether}.

As previously noted in subsection \ref{Sect: CylSym}, the gravitational case differs on a fundamental level from the electromagnetic one. The true acceleration (or force) and the true hidden momentum of each body (nearly) vanish in pure gravity. Furthermore, although the total momentum of the complete system is well-defined if the spacetime is asymptotically flat, there is, in general, no meaningful way of assigning one portion of this total to the gravitational field and another portion to the bodies. Field momentum need not be localizable or additive. Nevertheless, the analogy established in Sect. \ref{Sect: Analogy} suggests that in the 1.5PN approximation, each body acts {\it as though} it possessed a hidden momentum $\vec{S} \times \vec{a}$ (where $\vec{a} \equiv \rmd^2 \vec{z}/\rmd s^2$). This may be interpreted as arising from the usual kinematical contribution $-\vec{S} \times \vec{a}$ as well as a dynamical term $2 \vec{S} \times \vec{a}$ (corresponding to $\vec{\mu} \times \vec{E}$ in the electromagnetic case).  Furthermore, the system acts {\it as though}, during each orbit, the dynamical portion of the hidden momentum is exchanged with gravitational field momentum.

\subsection{Kicks}

We have shown in this paper that bobbing behavior is a ubiquitous feature of the motion of relativistic spinning bodies. Furthermore, we have shown explicitly that for the tetherball system, for the electromagnetic binary at 1.5PC order, and for the gravitational binary at 1.5PN order, the net bobbing is of a purely kinematic nature.

In the case of coalescing black holes, not only is bobbing behavior seen, but, in some cases, after merger the coalesced system receives a large ``velocity kick.'' In some cases, this kick appears to be a continuation of the net bobbing motion of the black holes just prior to merger. Thus, it appears natural to expect that the velocity kick is closely related to the bobbing behavior and that, indeed, these could  be two aspects of the same basic phenomenon. We now argue that this is not the case, and that the kicks and bobbing are essentially unrelated.

In the case of a merger of two bodies, the resulting body should move inertially and, by \eqref{MomVel}, would be expected to have negligible hidden momentum. Thus, in order have a velocity kick, it is necessary for the merged body to have a momentum kick as well. In order for this to be possible, it is necessary that momentum be carried off by another system. In the case of tetherballs, this is impossible, since there is no other system available. Thus, although the tetherball system bobs in a manner very similar to the black hole system, if the tetherballs were to merge, the merged system could not acquire any velocity kick.

On the other hand, a velocity kick should be possible in the merger of two charged spinning bodies. We have seen in the previous subsection that even in the 1.5PC approximation, a significant amount of momentum can be stored in the electromagnetic field (see \eqref{eq:Px}). If a merger event were to occur and if some\footnote{If we perform the integral in \eqref{moint} over a small ball enclosing body ${\rm B}$ rather than over all of space and then perform the same manipulations that led to \eqref{eq:Px}, we find that a surface term contribution of $\frac{1}{3} \vec{\mu}_\rmB \times \vec{E}_\rmA$ now arises from the integration by parts. We thereby find that the field momentum contained within the ball is $-\frac{2}{3} \vec{\mu}_\rmB \times \vec{E}_\rmA$. This calculation shows that $2/3$ of the momentum \eqref{eq:Px} stored in the electromagnetic field is located within---or very near---the bodies, and would not likely be released in a merger event.} of this field momentum could be ``released'' and allowed to radiate to infinity rather than be reabsorbed by the merged object, then the merged object would acquire a velocity kick.

The 1.5PC approximation is not adequate to analyze kicks, because at this level of approximation, the field momentum \eqref{eq:Px} can only be exchanged with the bodies and cannot be radiated to infinity.
However, it is not difficult to imagine that in a more relativistic system containing charged spinning bodies, a field momentum comparable to \eqref{eq:Px} could be radiated to infinity if a merger event were to occur.  Although many of the quantitative details would undoubtedly differ from the 1.5PC approximation, we would expect (net) bobbing to be governed primarily by spin and kicks to be governed primarily by magnetic dipole moment. In particular, if the charged bodies have spin but no magnetic dipole moment (e.g., they are composed partly of uncharged rotating matter and partly of charged non-rotating matter) then we would expect significant bobbing but no significant kick. By contrast, if the bodies have magnetic dipole moment but no spin (e.g., they contain oppositely charged portions that counter-rotate), then we would expect to have negligible net bobbing effects, but in suitable cases it might be possible for a merged object to acquire a large kick. Similarly, if in an electromagnetic system we were to reverse the magnetic dipole moments but not the spins, then we would expect the net bobbing to be the same but the kick to reverse sign.

In the gravitational case, both the kinematic net bobbing and the perpendicular field momentum are determined by $\vec{a} \times \vec{S}$ at 1.5PN order and cannot be varied independently. Thus, in the gravitational case, it should not be surprising if the perpendicular kick velocity is correlated with the phase of net bobbing at merger, and if the maximum perpendicular kick velocity is the same order of magnitude as the maximum net bobbing velocity.  Nevertheless, it is our view that bobbing and kicks are most naturally viewed as independent phenomena.
\begin{acknowledgments}

This research was supported in part by NSF Grants No. PHY04-56619
and No. PHY08-54807 to the University of Chicago.

\end{acknowledgments}

\end{document}